\documentclass[8.5pt,twoside,twocolumn]{article}
\usepackage[super,sort&compress,comma]{natbib} 
\usepackage{mhchem}
\usepackage{times,mathptm}
\usepackage{sectsty}
\usepackage{balance} 

\usepackage{graphicx} 
\usepackage{lastpage}
\usepackage[format=plain,justification=raggedright,singlelinecheck=false,font=small,labelfont=bf,labelsep=space]{caption} 
\usepackage{fancyhdr}
\usepackage{bm}
\usepackage{color}

\newcommand{\e}[1]{\times10^{#1}}
\newcommand{\gd}{\dot{\gamma}}

\begin{document}

\thispagestyle{plain}
\fancypagestyle{plain}{
\renewcommand{\headrulewidth}{1pt}}
\renewcommand{\thefootnote}{\fnsymbol{footnote}}
\renewcommand\footnoterule{\vspace*{1pt}%
\hrule width 3.4in height 0.4pt \vspace*{5pt}} 
\setcounter{secnumdepth}{5}

\makeatletter 
\def\subsubsection{\@startsection{subsubsection}{3}{10pt}{-1.25ex plus -1ex minus -.1ex}{0ex plus 0ex}{\normalsize\bf}} 
\def\paragraph{\@startsection{paragraph}{4}{10pt}{-1.25ex plus -1ex minus -.1ex}{0ex plus 0ex}{\normalsize\textit}} 
\renewcommand\@biblabel[1]{#1}            
\renewcommand\@makefntext[1]%
{\noindent\makebox[0pt][r]{\@thefnmark\,}#1}
\makeatother 
\renewcommand{\figurename}{\small{Fig.}~}
\sectionfont{\large}
\subsectionfont{\normalsize} 

\fancyfoot{}
\fancyfoot[RO]{\footnotesize{\sffamily{1--\pageref{LastPage} ~\textbar  \hspace{2pt}\thepage}}}
\fancyfoot[LE]{\footnotesize{\sffamily{\thepage~\textbar\hspace{3.45cm} 1--\pageref{LastPage}}}}
\fancyhead{}
\renewcommand{\headrulewidth}{1pt} 
\renewcommand{\footrulewidth}{1pt}

\setlength{\arrayrulewidth}{1pt}
\setlength{\columnsep}{6.5mm}
\setlength\bibsep{1pt}

\if{\thispagestyle{plain}
\fancypagestyle{plain}{
\renewcommand{\headrulewidth}{1pt}}
\renewcommand{\thefootnote}{\fnsymbol{footnote}}
\renewcommand\footnoterule{\vspace*{1pt}%
\hrule width 3.4in height 0.4pt \vspace*{5pt}} 
\setcounter{secnumdepth}{5}

\makeatletter 
\def\subsubsection{\@startsection{subsubsection}{3}{10pt}{-1.25ex plus -1ex minus -.1ex}{0ex plus 0ex}{\normalsize\bf}} 
\def\paragraph{\@startsection{paragraph}{4}{10pt}{-1.25ex plus -1ex minus -.1ex}{0ex plus 0ex}{\normalsize\textit}} 
\renewcommand\@biblabel[1]{#1}            
\renewcommand\@makefntext[1]%
{\noindent\makebox[0pt][r]{\@thefnmark\,}#1}
\makeatother 
\renewcommand{\figurename}{\small{Fig.}~}
\sectionfont{\large}
\subsectionfont{\normalsize} 

\fancyfoot{}
\fancyfoot[LO,RE]{\vspace{-7pt}\includegraphics[height=9pt]{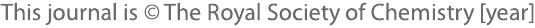}}
\fancyfoot[CO]{\vspace{-7.2pt}\hspace{12.2cm}\includegraphics{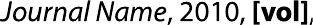}}
\fancyfoot[CE]{\vspace{-7.5pt}\hspace{-13.5cm}\includegraphics{RF.pdf}}
\fancyfoot[RO]{\footnotesize{\sffamily{1--\pageref{LastPage} ~\textbar  \hspace{2pt}\thepage}}}
\fancyfoot[LE]{\footnotesize{\sffamily{\thepage~\textbar\hspace{3.45cm} 1--\pageref{LastPage}}}}
\fancyhead{}
\renewcommand{\headrulewidth}{1pt} 
\renewcommand{\footrulewidth}{1pt}
\setlength{\arrayrulewidth}{1pt}
\setlength{\columnsep}{6.5mm}
\setlength\bibsep{1pt}
}\fi

\twocolumn[
  \begin{@twocolumnfalse}
\noindent\LARGE{\textbf{Rheology of Lamellar Liquid Crystals in Two and Three Dimensions: A Simulation Study}}
\vspace{0.6cm}

\noindent\large{\textbf{O. Henrich\textit{$^{a,c}$}, K. Stratford\textit{$^{b}$}, D. Marenduzzo\textit{$^{c}$}, P.~V. Coveney{$^{a}$} and M.~E. Cates \textit{$^{c}$}}
}\vspace{0.5cm}


\vspace{0.6cm}

\noindent \normalsize{We present large scale computer simulations of the nonlinear bulk rheology of lamellar phases (smectic liquid crystals) at moderate to large values of the shear rate (Peclet numbers $10-100$), in both two and three dimensions. In two dimensions we find that modest shear rates align the system and stabilise an almost regular lamellar phase, but high shear rates induce the nucleation and proliferation of defects, which in steady state is balanced by the annihilation of defects of opposite sign. The shear rate $\dot\gamma_c$ at onset of this second regime is controlled by thermodynamic and kinetic parameters; we offer a scaling analysis that relates $\dot\gamma_c$ to a critical ``capillary number'' involving those variables. Within the defect proliferation regime, the defects may be partially annealed by slowly decreasing the applied shear rate; this causes marked memory effects, and history-dependent rheology. Simulations in three dimensions show instead shear-induced ordering even at the highest shear rates studied here. This suggests that $\dot\gamma_c$ shifts markedly upward on increasing dimensionality. This may in part reflect the reduced constraints on defect motion, allowing them to find and annihilate each other more easily. Residual edge defects in the 3D aligned state mostly point along the flow velocity, an orientation impossible in two dimensions.}
\vspace{0.5cm}
 \end{@twocolumnfalse}
  ]


\footnotetext{\textit{$^{a}$ Centre for Computational Science, University College London, 20 Gordon Street, London WC1H 0AJ, UK}}
\footnotetext{\textit{$^{b}$ Edinburgh Parallel Computing Centre, School of Physics and Astronomy, University of Edinburgh, JCMB Kings Buildings, Mayfield Road, Edinburgh EH9 3JZ, UK}}
\footnotetext{\textit{$^{c}$ SUPA, School of Physics and Astronomy, University of Edinburgh, JCMB Kings Buildings, Mayfield Road, Edinburgh EH9 3JZ, UK}}



\section{Introduction}

Instances of soft complex fluids abound in our everyday lives: examples include paint (a colloidal suspension), mayonnaise (an emulsion), and engine oil (a polymer solution). Many soft materials, including the amphiphiles used in skin creams, fabric softeners, and some pharmaceutical products, form lamellar mesophases (also known as smectic liquid crystals) whose perfectly ordered state consists of a regular stack of flat layers. As well as these lyotropic liquid crystals arising in amphiphilic solutions, thermotropic smectics commonly arise in systems of rodlike, mesogenic molecules even without solvent~\cite{deGennes}. Such phases are of wide technological importance in devices
such as electronic displays and sensors.

The flow behaviour (rheology) of smectic liquid crystals affects many industrial and manufacturing processes; conversely, rheological measurements provide important probes of their macroscopic material properties.
For a simple fluid the response to steady applied stress is quantified in the linear (Newtonian) regime by a single viscosity. Lamellar phases are anisotropic, and thus in principle have several viscosities as well as an elastic modulus related to layer compression. However, these linear response parameters become somewhat irrelevant when defects are present: unless their density is very low, the defects soon dominate the stress. Moreover, since the defect pattern is strongly dependent on flow rate, the rheology is generically nonlinear. On shearing, the effective viscosity (ratio of shear stress $\sigma_{xy}$  to strain rate $\dot\gamma$) may either increase or decrease, corresponding
to shear thickening or shear thinning behaviour~\cite{Larson}. In addition, if a defect texture is present in the system at rest, this may show a viscosity divergence at small $\dot \gamma$ --- i.e., a yield stress.
Finally, in common with other complex fluids, lamellar phases can undergo transitions into heterogeneous states, for instance to form shear bands (layers of unequal $\sigma_{xy}$ or $\dot\gamma$) \cite{Olmsted08}. 

Lamellar liquid crystals provide an important paradigm for non-Newtonian fluids in which the microstructure of the material is strongly coupled to the flow. Crucially, the defect texture may depend on past history -- for instance, the strain rate in any prior flow that was suddenly switched off. This rheological hysteresis means that the precise flow protocol, including any pre-shearing applied prior to starting the experiment, needs to be specified. Such hysteresis can arise in polydomain liquid crystals of various types (not just smectics) such as nematics in which the existence and evolution of regions with misaligned directors creates rheological nonlinearity. A closely related example is the so-called `region I flow' in polymeric nematic liquid crystals, where it is thought that a defect network arises, with enhanced viscosity, but with
marked shear thinning as this structure is sheared out at high flow rates \cite{Larson}. 

An intriguing complication in addressing the dynamics of layered systems
is permeation,
where individual molecules in the fluid translate while keeping the
layerwise order intact. In steady state, permeation leads to increased dissipation
and hence to a large viscosity in any geometry where flow necessitates such motion \cite{deGennes}. (A well-known example of permeation flow
is found in cholesterics which are sheared or pushed along the direction
of the cholesteric helix.) Permeation is often involved in defect motion such as the growth or shrinkage of incomplete layers at the sites of edge dislocations, and may also be implicated in other forms of topological reconstruction. Such reconstruction arises under shear, sometimes to an extreme degree, for instance in the shear-induced transition from an ordered lamellar stack to an onion phase in lyotropic smectics \cite{Diat93,Panizza96,Iwashita07,Gradzielski03}. (The onion phase comprises a space-filling packing of multilayer vesicles; under some conditions similar defect textures can arise without shear \cite{Gomati87,Boltenhagen92,Fournier94,Ramos04}.) The mechanism underlying such global reconstructions under shear of the layer organization remains unclear. Most theoretical predictions have focussed on buckling and other instabilities of perfectly ordered smectics as the initial trigger for onion formation \cite{Zilman99,Courbin02,Guo2002,Soddemann2004, Stewart2009} but the pathway may also strongly involve the physics of defect motion \cite{Zipfel99, Leon00}. Although we do not see onions in the simulations reported here, in two dimensions we do see a transition from a well ordered phase to a defect-rich phase, whose possible link to onion formation deserves to be explored in future work. Another interesting avenue connects with the observation, in block copolymers and some lyotropic surfactant systems, of a shear-induced
phase transitions from a homogeneous state into ordered phases
like the lamellar one~\cite{Cates89,Koppi93,Fredrickson94}. However, theory suggests that thermal noise (causing local fluctuations about the mean state of order) plays a central role in this transition~\cite{Cates89}. In the present work we focus on the simplest case in which such thermal fluctuations are ignored, and the time evolution of the microstructure is deterministic (though certainly chaotic and unpredictable
once defects are present).

Despite a wealth of experimental results, and the importance of flow behaviour
to many commercial applications, mechanistic understanding of
the rheology of lamellar phases remains limited. This is largely
because of the nonlinear nature of the equations of motion, and the
coupling of the flow to different microstructures. These complications either require
drastic approximations or, as pursued here, use of computer
simulation. However, once defects are present, such simulations are computationally expensive. 
While sheared smectics have been studied extensively via simulations
in the literature~\cite{Swift96,Gonnella97,Gonnella98,Xu03,Xu05,Xu06a,Xu06b},
computational constraints have meant almost all these results are limited
to two dimensions. Even in 2D, the number of layers simulated has often not been large enough to meaningfully explore the dynamics of defect textures which can have a controlling influence on rheology. Moreover, for technical reasons a number of these studies have used boundary conditions (BCs) that are not optimal if one wants to predict the behaviour of bulk samples, for example using solid moving walls to drive the fluid flow rather than periodic BCs. This choice does not minimize finite size effects and, when combined with lattice Boltzmann algorithms, severely limits the shear rates attainable in large systems. Moreover, solid walls for the fluid velocity have sometimes been used in conjunction with periodic BCs for the order parameter \cite{Xu06b} -- a somewhat inconsistent combination that makes it difficult to disentangle bulk and boundary effects. 

In this work, we try to overcome these technical obstacles, presenting systematic results for large two dimensional simulations with fully periodic boundary conditions.
Guided by those results, we additionally report some very large three dimensional simulations at selected points in the parameter space. These 3D simulations are, we believe, of unprecedented scale for lamellar phases, although even larger ones have recently been used to address the rheology of cubic liquid crystals \cite{Saksena09a, Saksena09b}.

Below, we describe smectics via a widely-used free energy functional \cite{Gompper, Gonnella97, Xu06b, Cencini07}. This provides a robust generic framework for addressing lamellar ordering without being tied to any specific class of materials (although we do become more specific when estimating suitable parameters for the simulations). The free energy is formulated in terms of a scalar order parameter $\phi({\bf r})$. This can be viewed as describing a composition variable in a binary fluid system stabilized by added amphiphiles; one can then speak of alternating layers of fluid, with interfaces between these layers. A similar interpretation applies to block copolymers where the composition variable describes the relative local density of the two blocks. However, the same order parameter could equally describe the local number density of the midpoints of a rodlike mesogenic molecule (measured relative to the mean) which can develop a periodic density wave that represents a thermotropic smectic state.

The governing equations in our model are a convection-diffusion equation
for the scalar order parameter, and the Navier-Stokes equation describing
the hydrodynamics of fluid flow. The stress tensor for the fluid contains not only a Newtonian viscous term but extra contributions arising from local deviations of the order parameter from its state of least free energy. Diffusive relaxation of the order parameter then causes additional dissipation, resulting in an effective viscosity that can be much higher than the intrinsic Newtonian contribution.
Our dynamical equations are solved using a combination of a finite
difference scheme for the convection-diffusion equation, and the
lattice Boltzmann method (LBM)~\cite{Succi} for the Navier-Stokes equation \cite{Marenduzzo07, Henrich10a,  Henrich10b, Henrich11a,Gonnella97}.
We use sliding periodic (Lees-Edwards) boundary conditions for
both the convection-diffusion and the Navier-Stokes equations. Note that while these boundary conditions impose a fixed integrated strain rate across the whole system they do not impose a uniform strain rate locally; shear bands can therefore arise if the system so chooses. (As it happens, we do not observe shear bands in the range of shear rates we address here.)

Our main finding is that, in two dimensions at least, sheared smectics can show two completely different types of rheological response. At modest rates
of shear, where thermodynamic forces are relatively strong compared with
flow effects,  shear quickly aligns the system into the perfect lamellar phase preferred by thermodynamics,
with only a few defects. The rheology in this regime is quasi-Newtonian, with shear stress linear in the applied shear rate. (There is however a large first normal stress difference.) At higher shear rates, larger thermodynamic forces are generated as layers fail to keep up with the imposed flow. This leads to the nucleation of many new defects, balanced in steady state by annihilation when defects of opposite sign encounter each other. 

The defects do not appear as a direct result of spontaneous thermal fluctuations. Instead they are a result of an undulation instability in our (deterministic) nonlinear dynamics. This occurs when viscous forces begin to dominate interfacial forces causing layers to pinch off and break, creating new defects in the process. The onset of a nontrivial topological pattern under deterministic evolution equations in a complex fluid is similar to the case of sheared binary fluids undergoing phase separation \cite{Stratford07}.

The strain rate $\dot\gamma_c$ needed to enter this nonlinear flow regime depends on thermodynamic and kinetic parameters. A comprehensive exploration of parameter space is not possible, but we give a scaling argument which relates $\dot\gamma_c$ to those parameters through an effective capillary number, formed as the ratio of viscous to thermodynamic stress.
Within the defect proliferation regime, the macroscopic rheology is generically nonlinear, and shows shear thinning. The shear stress appears to vary as a sublinear power law, $\sigma_{xy}\propto\dot{\gamma}^m$, with $m\simeq 0.87$ for typical parameter values, although the dependence of $m$ on the parameters precludes interpretation of this as a universal scaling law.
Such a behaviour is qualitatively reminiscent of the one found experimentally in~\cite{Lu08}, where the numerical value of $m$ was however different. In this strongly non-Newtonian regime we also find that the rheology is history dependent. 

Finally, and intriguingly, we find that dimensionality appears to play a key role. Within our 3D simulations, even when performed with parameter values identical to those used in 2D, we have so far not been able to enter the nonlinear flow regime associated with defect proliferation. Instead, we always observe shear-induced ordering into a lamellar phase with quasi-Newtonian rheology. Further research, requiring careful and systematic parameter steering, will be needed to map out quantitatively the onset of the defect proliferation regime, if one exists, in three dimensions.
 
This paper is organized as follows. In Section \ref{themodel} we introduce the model,
including its equilibrium properties, the full stress tensor and the
dynamical equations. We briefly sketch how we derive structural information
from the order parameter and describe a method for quantitative analysis of the defect density. Section \ref{parametermap} describes the mapping between our simulation parameters and those for physical systems of interest.
We present results for the two-dimensional case in Section \ref{results2D}, discussing
the morphology as a function of flow rate. We also consider average and local shear stress densities and
first normal stress differences, and explore their relation to the defect structure. In addition we report Fourier-space information (structure factors) under flow.
In Section \ref{results3D} we present and compare our results for 3D simulations which, by computational necessity, are less comprehensive than in 2D. 
A summary with conclusions closes the paper in Section \ref{conclusions}.

\section{Model and Definitions} \label{themodel}

The equilibrium phase in a system of volume $V$ is found for an order parameter field $\phi({\bm r})\equiv \phi$ that minimises the Landau-Brazovskii free energy functional \cite{Brazovskii75, Gompper}.
\begin{eqnarray}
{\cal F}[\phi] &=& \int d^dx f(\phi,\nabla\phi,\nabla^2\phi) = V\bar f
\label{FE-functional}
\\ &=& \int d^dx\left\{\frac{a}{2}\phi^2+\frac{b}{4}\phi^4+\frac{\kappa}{2} ({\nabla}\phi)^2+\frac{c}{2}(\nabla^2\phi)^2\right\}.
\nonumber
\end{eqnarray}
In minimizing this free energy at equilibrium, we ignore thermal fluctuations (treated in \cite{Brazovskii75,Cates89}), thereby adopting a mean-field approximation as in previous literature \cite{Kendon01,Gonnella97, Gonnella98, Xu03, Xu05, Xu06a, Xu06b}. This should be a good approximation far from the ordering transition at which $\phi$ first becomes nonuniform. The uniform state has $\phi = 0$ and hence ${\cal F}= 0$.

The parameter range relevant to lamellar phases has $\kappa < 0$ and $c>0$. Then, if we suppress the quartic term, the free energy density of a sinusoidal lamellar state $\phi({\bf r}) = \phi_k\cos({\bf k}.{\bf r})$ obeys $\bar f = (a+\kappa k^2+ c k^4)\phi_k^2/4$. This is minimized by choosing $k=q\equiv\sqrt{|\kappa|/2c}$; the free energy density is then $\bar f = (a-\kappa^2/4c)\phi_q^2/4$ so that ordering will occur for $a<a_c=\kappa^2/4c$. The amplitude $\phi_q$ is then unbounded without the (positive) quartic term, since $\bar f$ can in that manner become infinitely negative. Restoring the quartic term now causes saturation at $\phi_q\sim \sqrt{4(a_c-a)/3b}$, creating also nonlinear harmonics in the density profile. The final free energy density obeys $\bar f \simeq (a_c-a)^2/b$. By similar arguments, the characteristic elastic modulus associated with microstructural deformations is
\begin{equation}
\bar g\sim \frac{(a_c-a)a_c}{4b}
\label{barg}
\end{equation}
which is the energy density of the gradient terms only. The terms involving $\phi^2$ and $\phi^4$ in Eq.~\ref{FE-functional} are unaltered by affine deformations which conserve density locally; hence those terms do not contribute to $\bar g$.

The dynamics of the order parameter is governed by an advection-diffusion equation with a diffusive current of $\phi$ dependent on the chemical potential $\mu$, the functional derivative of ${\cal F}$ with respect to $\phi$, as follows:
\begin{equation}\label{OPEOM}
\partial_t \phi +{\nabla}({\bm u}\phi) = M \nabla^2 \frac{\delta \cal F}{\delta \phi}=M \nabla^2 \mu.
\end{equation}
Here $M$ is a mobility coefficient. A wavevector-dependent linearized collective diffusivity $D_k$ can then be defined via $\dot\phi_k = -k^2D_k\phi_k$, whose value at the ordering wavevector obeys $D_q = M(a-a_c)/2$ (this is negative, signifying the instability). In contrast, an estimate of the molecular diffusivity can be made by considering zero wavevector, setting $b = -a$ (see below) and examining the relaxation equation for small deviations around the uniformly ordered state at $\phi = \pm 1$. This gives $\dot\phi = 2bM\nabla^2\phi$ so that the molecular diffusivity is of order $\tilde D = 2Mb$. 

Treating the fluid as incompressible, the evolution of the velocity ${\bm u}$ obeys the Navier-Stokes equation
\begin{equation}\label{NSE}
\rho(\partial_t u_\alpha + (u_\beta \nabla_\beta) u_\alpha) = \partial_\beta \sigma_{\alpha \beta},
\end{equation}
where $\rho$ is the mass density. The stress tensor in this equation reads
\begin{equation}
\sigma_{\alpha \beta}=\eta (\partial_\alpha u_\beta + \partial_\beta u_\alpha)-p\delta_{\alpha\beta} - P_{\alpha \beta}. \label{fullstress}
\end{equation}
This consists of a Newtonian hydrodynamic part, which is proportional to the background viscosity $\eta$ and the symmetrized velocity gradient, a hydrostatic pressure term, which includes an isotropic stress contribution from the order parameter 
\begin{eqnarray}
p_{\phi}&=& \frac{a}{2}\phi^2+\frac{3b}{4}\phi^4-\kappa\left(\phi(\nabla^2\phi)+\frac{1}{2}({\nabla}\phi)^2\right)\nonumber\\
& &\quad+c\left(\phi(\nabla^2)^2\phi+\frac{1}{2}(\nabla^2\phi)^2+\partial_{\gamma}\phi\partial_\gamma(\nabla^2\phi)\right)
\end{eqnarray}
and finally another term consisting of the order-parameter contribution to the deviatoric pressure~\cite{Swift96,Yang76,Evans79} 
\begin{equation}
P_{\alpha \beta}=\kappa\partial_\alpha\phi\partial_\beta\phi -c \left(\partial_\alpha\phi\partial_\beta(\nabla^2\phi)+\partial_\beta\phi\partial_\alpha(\nabla^2\phi)\right).
\end{equation}

In our hybrid LB scheme, the Navier-Stokes equation (\ref{NSE}) is solved at every timestep by means of a multi-relaxation time lattice-Boltzmann solver \cite{dHumieres02,Adhikari05} using the D3Q19 model.
The velocity field is then used in the advection-diffusion equation Eq.\ref{OPEOM}, which is solved by a finite-difference method at the same timestep. We assume periodic boundary conditions along the flow and vorticity directions and Lees-Edwards sliding periodic boundary conditions\cite{leesedwards,Wagner02} along the direction of the velocity gradient.
Following \cite{Wagner02}, we introduce
additional internal Lees-Edwards planes to overcome an upper limit on shear rate that would otherwise be inversely proportional to the system size in the gradient direction. (This restriction arises because fluid velocities must be small compared to the sound speed in the frame of the lattice; the additional planes create a stack of lattice slabs whose velocity is nearly matched to the local fluid velocity.) 
The total number of Lees-Edwards planes was varied according to the size of the simulation so that the separation of two planes was always 16 lattice sites.
We relegate further comments on the implementation of the Lees-Edwards boundary
conditions to an Appendix. 

Structural information in Fourier space was obtained from the static structure factor $C({\mathbf k})$, defined as
\begin{equation}
C({\mathbf k})=\phi({\mathbf k})\phi(-{\mathbf k}),
\label{S}\end{equation}
with $\phi({\mathbf k})$ the normalized Fourier transform of the order parameter.
The transform itself was calculated using the FFTW-library \cite{FFTW}.
In principle an ensemble-average or time-average should appear in Eq.~\ref{S} so that $C$ becomes a smooth function of wavevector; in practice an instantaneous measurement, plotted as a colour intensity map, was found sufficient.

For the analysis of the relation between rheological quantities and the morphology of the system we need to quantify the number of defects.
For this purpose we defined an alignment function $X({\mathbf r})$, which measures the average orientation of the order parameter gradients, as follows
\begin{equation}\label{alignment}
X({\mathbf r})=\frac{1}{V'({\mathbf r})} \int_{V'({\mathbf r})} d{\mathbf r'} \left(\frac{\nabla \phi ({\mathbf r}) \cdot \nabla \phi ({\mathbf r'})}{|\nabla\phi ({\mathbf r}) ||\nabla\phi ({\mathbf r'}) |}\right)^2. 
\end{equation}
Here $V'$ is a small volume that contains $\mathbf r$; in practice $V'$ can be chosen to contain $\mathbf r$ and its nearest neighbours only.

For ordered regions with small curvature of the lamellar interfaces, $X$ is very close to unity. Wherever lamellae terminate at edge dislocations, and in regions where their orientation changes abruptly, $X$ departs from unity, taking smaller positive values.
By setting a threshold $X=\xi$ it is possible to detect regions where the order parameter structure deviates significantly from what one would recognize visually as an aligned lamellar configuration.
We found a threshold value of $\xi=0.8$ to be well suited to our requirements. This protocol does not directly count the number of defects, since at each defect more than one site has an alignment function below the threshold.
However, as the number of sub-threshold sites per defect is reasonably constant, the degree of order under flow can usefully be quantified by defining a ``defect density" as the system-wide fraction of sub-threshold sites:
\begin{equation}\label{defect density}
\rho_D({\mathbf r})=\frac{1}{V} \int d{\mathbf r'} \theta(\xi-X(\mathbf r')).  
\end{equation}
Here $\theta$ is the Heaviside function. The conversion factor between $\rho_D$ and the defect number density (in 2D) or line density (in 3D) depends on the lamellar wavenumber $q$ and the choice of threshold parameter $\xi$; we do not make such conversions here and consider $\rho_D$ itself as an appropriate measure of disorder.

\section{Parameter Mapping to Physical Units}
\label{parametermap}

The Landau-Brazovskii model belongs to the class of phenomenological models of liquid-crystalline materials.
This means it features an isotropic-to-lamellar phase transition without explaining the underlying microscopic mechanism.
Hence there is a many-to-one mapping between physical parameters and those of the phenomenological model. However, in the context of any particular system, it is possible to indicate physical parameters for which the chosen model is a good representation, and we do this below.

The parameter space of the model is quite large. Although in exploratory work we performed studies across a wider range of parameter values, for simplicity in the results presented here we restrict attention to fixed values of $\kappa = -6\times 10^{-4}$LBU and $c = 7.6\times 10^{-4}$LBU (here LBU denotes LB units as defined below). These are chosen to give a lamellar wavelength of $L = 10$ LBU (that is, $q = \sqrt{|\kappa|/2c} = 2\pi/10$LBU) which offers suitable spatial resolution. Moreover the microphase separation point, set by $a_c = \kappa^2/4c = 1.2\times 10^{-4}$LBU, is then a fixed constant. 

As usual we set the lattice spacing $l$, time step $\Delta t$ and mass density of the fluid $\rho$ all to unity to define LB units (LBU). We need to calibrate these units (or equivalently, length, energy density and time) against real physical units. We do this first for thermotropic smectics, then consider the lyotropic case.

To calibrate the LB length unit, we equate $L$ the lamellar wavelength, with that of a typical thermotropic smectic: $10$ LBU $= L = 4$nm. Hence the LB length unit is about $0.4$nm; and 1m$= 2.5\times 10^{9}$LBU. Free energy density parameters such as $a_c \simeq 10^{-4}$LBU should then be of order $\alpha^2k_BT/L^3$ where $\alpha =10$ is a molecular aspect ratio. Thus the LB unit of energy density is about $6\times 10^{10}$Pa, i.e. 1 Pa$= 1.6 \times 10^{-11}$ LBU.  

Turning now to the fluid viscosity $\eta$, after pilot runs spanning a range of several decades, we selected $8.33\times 10^{-2}$LBU for the production runs reported here. This represents the baseline viscosity of the isotropic phase close to the transition (not the higher, effective viscosity of the smectic once formed). Requiring this to be about ten times the viscosity of water (which is $\eta_w = 8.9\times 10^{-4}$Pa s) we find that the LBU unit of viscosity is about $0.1$Pa s. Given the previous result for the energy density, we infer that 1s$= 6\times 10^{11}$ LBU. The LBU timestep is thus around $1.6\times 10^{-12}$s. 

As detailed previously, an estimate of the molecular diffusivity is $\tilde D = 2Mb$; we set $M = 0.25$LBU throughout this study, so that $\tilde D$ lies in the range $2.5\times 10^{-5} - 2.5\times 10^{-4}$LBU. Taking a mid-range value of  $\tilde D = 10^{-4}$LBU and using the above conversion factors (1 m$= 2.5\times 10^9$LBU, 1s $= 6\times 10^{11}$LBU) gives $D\simeq 10^{-11}$m$^2$s$^{-1}$ which is a reasonable mesogenic value.

Recall that the mass density of our fluid is unity in LBU. However, combining the above conversion factors for length,  energy density, and time shows the physical and simulation units of mass density to be related via 1 kg m$^{-3} \equiv 1$ Pa s$^2$ m$^{-2} \simeq 10^{-8}$LBU. Thus our simulations address a fluid of mass density $10^6$kg m$^{-3}$ which is $10^3$ times larger than the true value for a typical material. The reason for this is that the LB algorithm uses fluid inertia to update the dynamics and the discretization should therefore involve replacing the real fluid with the densest one possible, while ensuring that the Reynolds number $Re \sim \rho\dot\gamma L^2/\eta$ (which for real smectics is exceedingly small) remains small enough for the simulation to faithfully describe the low $Re$ limit. As discussed elsewhere \cite{Cates04, Cates09} $Re$ values up to about 0.1 are then admissable; our simulations comply with this requirement. 

Finally, our shear rates lie in the range 
$10^{-5}-10^{-4}$LBU, so that the (dimensionless) Peclet number, $Pe =\dot\gamma L^2/\tilde D$, which expresses the relative importance of advective to diffusive transport, lies in the range $5-800$. These are moderate to high values where significant structural rearrangement can be expected. However, for thermotropic smectics, given our time unit of 1LBU$=1.6\times 10^{-12}$s, they correspond to strain rates of order $10^6$s$^{-1}$. This lies well above the range typically addressable with rheological experiments, but below that recently given in a theoretical study \cite{Stewart2009}.

The shear rates needed to achieve this range of $Pe$ fall dramatically when lyotropic smectics are considered. For lyotropics one has $L\sim 40$nm; the LB length unit is thus about 4nm. The factor $\alpha$ in the free energy density is now of order unity (since the main contributions are entropic and for flexible membranes there is about one relevant degree of freedom per cube of side $L$)~\cite{Safran}. The LB unit of energy density is then about $6\times 10^{5}$Pa, i.e. 1 Pa$= 1.6 \times 10^{-6}$ LBU. Both the viscosity $\eta$ and the molecular diffusivity $\tilde D$ remain comparable (within one order of magnitude) to the thermotropic values; from the viscosity we then determine that 1s$=6\times 10^6$LBU or 1LB timestep $=1.6\times 10^{-7}$s. Accordingly the strain rates relevant to the $Pe$ range addressed here are of order $10^2$ s$^{-1}$ and well within experimental reach, but below that addressed in molecular dynamics simulations \cite{Guo2002,Soddemann2004}. Note that for lyotropics we are effectively simulating a fluid of even higher mass density than for thermotropics; but since the simulated Reynolds numbers are unaffected by the change in parameter mapping, they are still sufficiently small~\cite{Cates04}. 

To further simplify our parameter exploration, throughout what follows we restrict attention to  states with $b = -a$. This convention has been widely used to study bulk spinodal decomposition in binary fluids where it can be absorbed in a rescaling of the order parameter \cite{Kendon01}: put differently, $b=-a$ ensures that coexisting bulk ordered phases have $\phi = \pm\phi_0$ with $\phi_0=1$. In lamellar states of $b= -a$ the amplitude of the lamellar ordering need not be unity but varies like $\phi_q \sim \sqrt{1+a_c/b}$ as stated previously. Thus the chosen value of $b$ can be thought of as controlling whether the peak order parameter variation is close to that arising for bulk coexistence (when $b\gg a_c$), or whether the amplitude $\phi_q$ is instead much larger ($b\ll a_c$) and determined by a competition between the gradient terms and a relatively weak quartic nonlinearity.  In this study we choose three values, $b = 5\times 10^{-4}, 1\times 10^{-4}$ and $5\times 10^{-5}$LBU so that the dimensionless quantity $\beta \equiv b/a_c$ is respectively $4.2,0.83,0.42$, lying near the crossover between these regimes. The observed $\phi_q$ range is about $1.2-1.5$, somewhat narrower than the range $1.1-1.8$ predicted from $\phi_q \sim \sqrt{1+a_c/b}$. However the latter estimate, which does not treat the nonlinear term exactly, is only a guideline.

Alongside the Reynolds number (which we deem irrelevant), $\beta$, and the Peclet number, we can identify two further dimensionless numbers of interest. One is the Schmidt number $Sc = \eta/\rho\tilde D$, which is large when momentum can diffuse much more rapidly than molecules - as applies in almost all dense fluids.  In our simulations $Sc$ is of order $100-1000$ which is safely large. (The work of Kumaran \cite{Kumaran2011} instead addresses Schmidt numbers in the range $0.2-5$ where different physics may apply.) The second parameter controls the strength of feedback between distortions of the $\phi$ field and the fluid motion. This 
can be expressed in terms of a characteristic value of the Peclet number, $Pe^*$, at which the free energy density in the order parameter field that couples to deformation (Eq.\ref{barg}), $\bar g=(a_c+b)a_c/4b$, is comparable to the viscous stress $\sigma_v=\eta\dot\gamma = (\eta \tilde D/L^2)Pe$. Equating these gives
\begin{equation}
Pe^* =\frac{(a_c+b)a_c}{8b^2}\,\frac{L^2}{\eta M}.
\label{pestar}
\end{equation}
With our chosen parameters (in LBU, $a_c=1.2\times 10^{-4}, L = 10, \eta = 8.3\times 10^{-2}, M = 0.25$) we find $Pe^* \sim 180$ at $\beta = 4.2$ and $Pe^* \sim 5000$ at $\beta = 0.42$. The maximum values of $Pe/Pe^*$ that we address below are of order 0.2 and 0.08 respectively in the two cases.

For later convenience we may define a new dimensionless number,
which we can call a capillary number. (The capillary number is conventionally defined as the ratio of viscous to interfacial forces in binary fluid flow.) We define this as:
\begin{equation}
Ca = \frac{Pe}{Pe^*} = \frac{4\eta\dot\gamma b}{(a_c+b)a_c}
\label{capillary}
\end{equation}
We will argue below that this could be a more relevant choice than $Pe$ for understanding the dependence of morphology on flow rate in smectic systems, at least when $Pe^*\gg 1$.
\vfill

\section{Results and Discussion}

We first present detailed results for two parameter sets ($\beta = 4.2, 0.42$) in two dimensions which we call systems A-2D and B-2D. System lattice sizes and shear rates are reported in Table \ref{tab1}; all remaining parameters are the same in both systems as detailed above. 
A third parameter set ($\beta = 0.83$, system C-2D) gave results broadly similar to B-2D. We will later report simulations in three dimensions for systems A-3D and C-3D which share parameters with A-2D and C-2D (see Table \ref{tab1}).
Note that our simulations in two dimensions are equivalent to assuming, in three dimensions, that the system maintains translational invariance in the vorticity direction. We find this is an increasingly good approximation at very low shear rates.

\begin{table}[htp]
\small
\begin{tabular*}{0.52\textwidth}{l|l|l|l|l|l|l}
\hline
run \# & $N_x\times N_y\times N_z$ & time $/10^5$ & $10^5b$ & Ca & Pe & $10^5\dot\gamma$\\
\hline
A & & & & & &\\
\hline 
1 & $1024\times512\times1$ & 0-3.2 & $50$ &0 &0 & 0 \\
2 & $1024\times512\times1$ & 3.2-6.4& $50$ & 0.224 & 40 &  $10$\\
3 & $1024\times512\times1$ & 6.4-9.6& $50$ &0 & 0 &  0 \\
\hline
4 & $512\times256\times1$ & 0-3.2 & $50$ &0 & 0 &  0\\
5 & $512\times256\times1$ & 3.2-6.4& $50$ &0.448 & 80 &  $20$ \\
6 & $512\times256\times1$ & 3.2-25.6& $50$ &0.224 & 40 &  $10$ \\
7 & $512\times256\times1$ & 12.8-38.4& $50$ &0.112 & 20 &  $ 5$ \\
8 & $512\times256\times1$ & 51.2-76.8& $50$ &0.056 & 10 &  $ 2.5$ \\
9 & $512\times256\times1$ & 51.2-102.4& $50$ &0.028 & 5 &  $1.25$ \\
\hline
10 & $256\times128\times1$ & 3.2-16.0& $50$ &0.22 & 40 &  $10$ \\
\hline
11 & $256\times128\times128$ & 0-6.4 & $50$ &0 & 0 &  0\\
12 & $256\times128\times128$ & 3.2-6.4 & $50$ &0.448 & 80 &  $20$\\
13 & $256\times128\times128$ & 3.2-9.6 & $50$ &0.224 & 40 &  $10$\\
14 & $256\times128\times128$ & 6.4-12.8 & $50$ &0.112 & 20 &  $5$\\
\hline
B & & & & & &\\
\hline
15 & $1024\times512\times1$ & 0-3.2 & $5$ &0 & 0 &  0 \\
16 & $1024\times512\times1$ & 3.2-9.6& $5$ &0.082 & 400 &  $10$ \\
17 & $1024\times512\times1$ & 9.6-12.8& $5$ &0 & 0 &  0 \\
\hline
18 & $512\times256\times1$ & 0-3.2 & $5$ & & 0 &  0 \\
19 & $512\times256\times1$ & 3.2-9.6& $5$ &0.164 & 800 &  $20$ \\
20 & $512\times256\times1$ & 3.2-25.6& $5$ &0.082 & 400 &  $10$ \\
21 & $512\times256\times1$ & 12.8-38.4& $5$ &0.041 & 200 &  $5$ \\
22 & $512\times256\times1$ & 25.6-64.0& $5$ &0.020 & 100 &  $2.5$ \\
23 & $512\times256\times1$ & 51.2-76.8& $5$ &0.010 & 50 &  $1.25$ \\
\hline
C & & & & & &\\
\hline
24 & $256\times128\times1$ & 0-3.2& $10$ &0 & 0 & 0 \\
25 & $256\times128\times1$ & 3.2-9.6& $10$ &0.252 & 400 &  $20$ \\
26 & $256\times128\times1$ & 3.2-12.8& $10$ &0.126 & 200 &  $10$ \\
\hline
27 & $256\times128\times128$ & 0-3.2 & $10$ &0 & 0 &  0 \\
28 & $256\times128\times128$ & 3.2-9.6 & $10$ &0.252 & 400 &  $20$\\
29 & $256\times128\times128$ & 3.2-12.8 & $10$ &0.126 & 200 &  $10$\\
\hline
\end{tabular*}

\caption{Simulation parameters: After initialization from a noisy uniform state, the systems were first equilibrated at zero shear rate, followed by a sequence of various shear rates, ranging from $\dot{\gamma}=1.25\times10^{-5}$ to $2\times10^{-4}$LBU. In some cases we performed a final switch off run at zero shear rate. The other model parameters $\kappa=-6\times10^{-4}, c=7.6\times10^{-4}, M=0.25$ and $\eta=8.33\times10^{-2}$ were the same throughout all simulations.}
\label{tab1}
\end{table}

\subsection{Two Dimensions}

\label{results2D}
Before presenting specific results for the parameter sets detailed above, we briefly describe the effect of wider parameter variations.
Within the Landau-Brazovskii free energy functional (Eq.~\ref{FE-functional}) a range of parameter variations can produce stable lamellar phases, as explored in previous studies \cite{Kendon01,Kumaran2001,Xu03, Xu06b,Kumaran2011}.
The nonequilibrium morphology of the lamellar structure, and the phase ordering kinetics (starting from a slightly noisy, uniform initial state) do however depend on the choice made for these parameters, and also for the viscosity. 
Formation and ripening of the lamellar structure is slower when $\beta$ is small, consistent with the reduced initial driving force for ordering in this case. Higher viscosities lead to more locally disordered structures in which there is a shorter separation between kinks or edge defects as one moves along the lamellae; conversely lower viscosities typically yield ``longer" lamellae.
This effect has been reported before \cite{Gonnella97}; we found it to be more pronounced for softer (lower $\beta$) systems. At fixed $\dot\gamma$, higher values of $\eta$ were found to cause greater inhomogeneity of lamellar shape and thickness, consistent with the fact that $Ca\propto\eta$ (the larger the shear stress, the larger the lamellar deformation). 

\begin{figure}[htp]
\centering
\includegraphics[angle=0,width=0.5\textwidth]{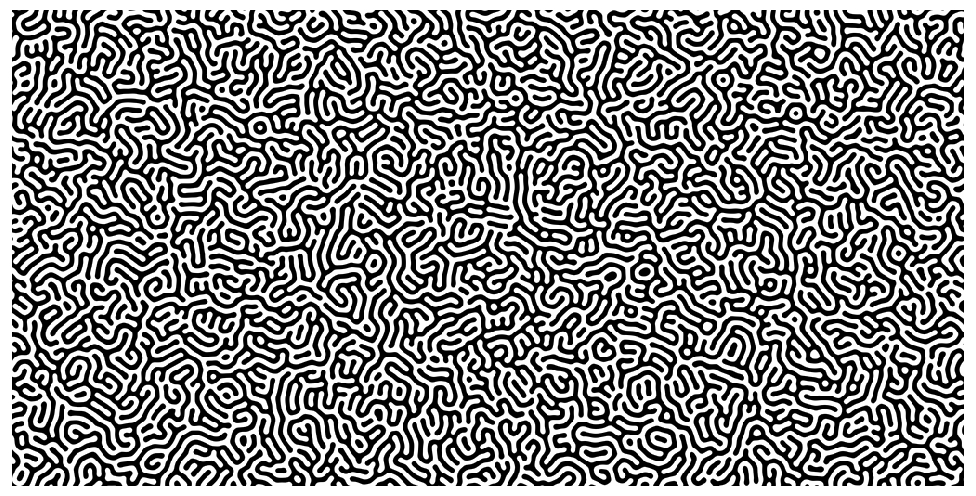}\\
\includegraphics[angle=0,width=0.5\textwidth]{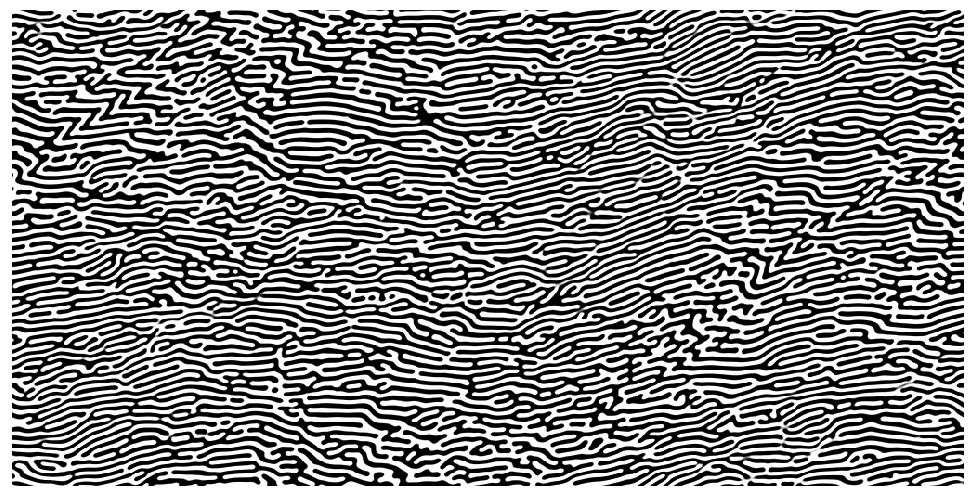}\\
\includegraphics[angle=0,width=0.5\textwidth]{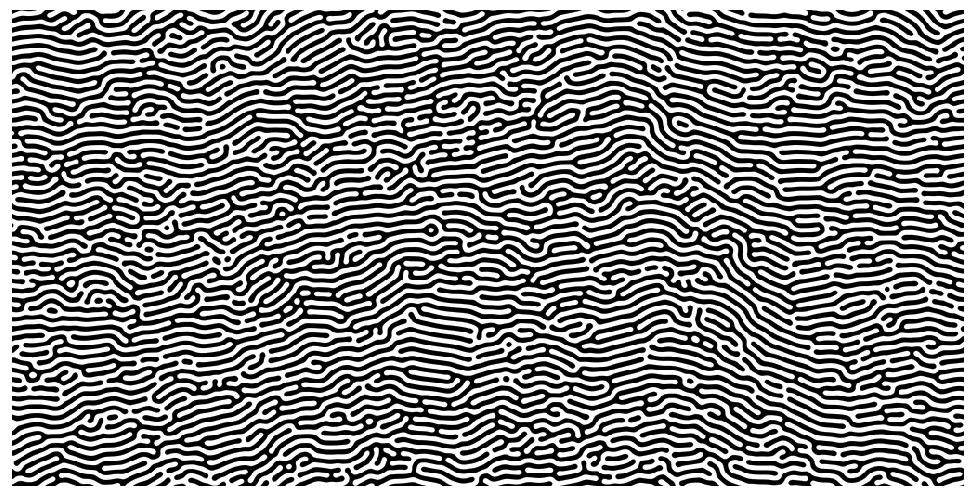}
\caption{Order parameter pattern $\phi({\bf r})$ of system A-2D (thresholded to enhance visibility): after equilibration (timestep $t=3.2\e{5}$, top), in steady shear at $\dot{\gamma}=1\times10^{-4}$ ($t=5\e{5}$, centre) and after switch-off ($t=1.28\e{6}$, bottom). Gray scaling from black to white corresponds to values of $-1\le\phi\le1$.}
\label{fig1}
\end{figure}

We turn now to system A-2D ($\beta = 4.2$). The top panel in Fig. \ref{fig1} shows the order parameter pattern $\phi$ at the end of the equilibration phase prior to shearing.
Although the true equilibrium structure would show homogeneous alignment, this cannot be reached in a reasonable time scale unless shear is applied. The resulting metastable structure is amorphous and isotropic, but locally exhibits the same lamellar spacing $L = 10$ LBU as the fully ordered equilibrium state.
If shear flow is imposed, the lamellae start to align in the flow direction, which induces a morphological changes to the structure. A typical snapshot in the resulting steady state at $Pe = 40$ $(Ca = 0.22)$ is shown in the middle panel of Fig.~\ref{fig1}. In this and all subsequent images, the applied flow velocity is horizontal and the velocity gradient vertical, with the flow directed to the right in the upper half and to the left in the lower half of the picture. After a short transient during which the order parameter pattern remains relatively homogenous (not shown) there emerge extended regions in which the smectic layers are either expanded or compressed, so that $L$ locally deviates from the equilibrium value. These mesoscopic inhomogeneities are inclined at an angle to the flow and remain continuously present in the steady state, but are not stationary: their local features are advected by the flow and evolve continuously by deformation. 

We found that after the shear flow was switched off the lamellar spacing returned to its equilibrium value everywhere, with expanded regions contracting and vice versa. The final state, after all relaxation has (virtually) come to a standstill, is shown in the bottom panel of Fig.~\ref{fig1}. 
Most of the lamellae remain oriented along the flow direction, but depending on the local expansion and compression undergone during the previous flow, some regions retain strong misalignment which does not anneal away. The separation between defects as one moves along the lamellae varies considerably in space; both short and long separations can be seen. This applies both to the aligned layers and to layers that are misaligned with the flow. 

\begin{figure}[htp]
\centering
\includegraphics[angle=0,width=0.5\textwidth]{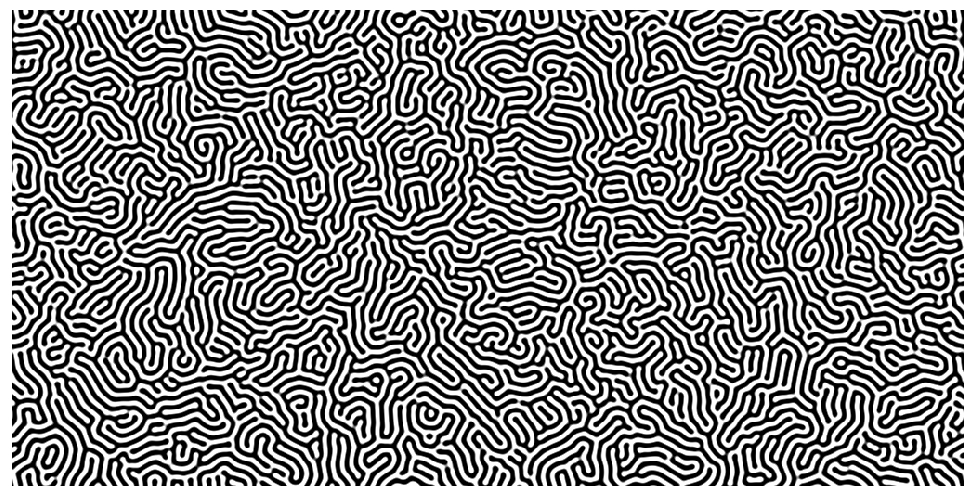}\\
\includegraphics[angle=0,width=0.5\textwidth]{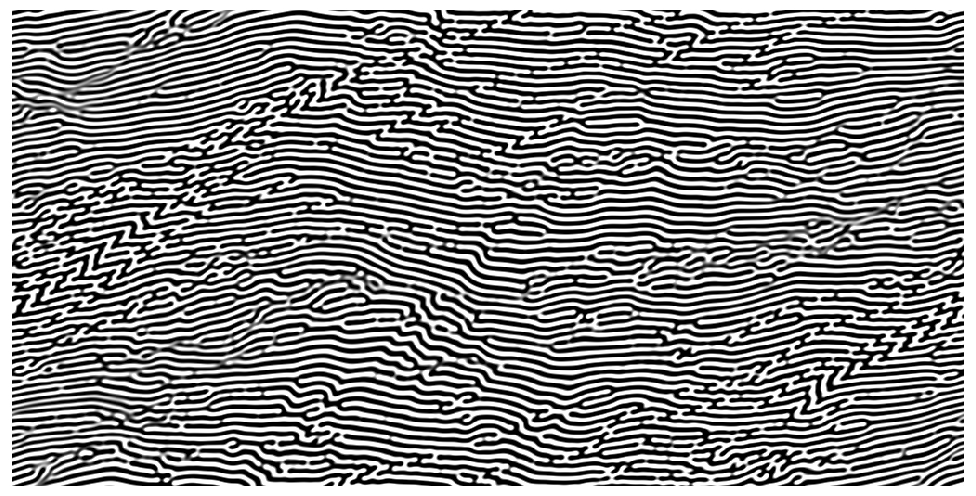}\\
\includegraphics[angle=0,width=0.5\textwidth]{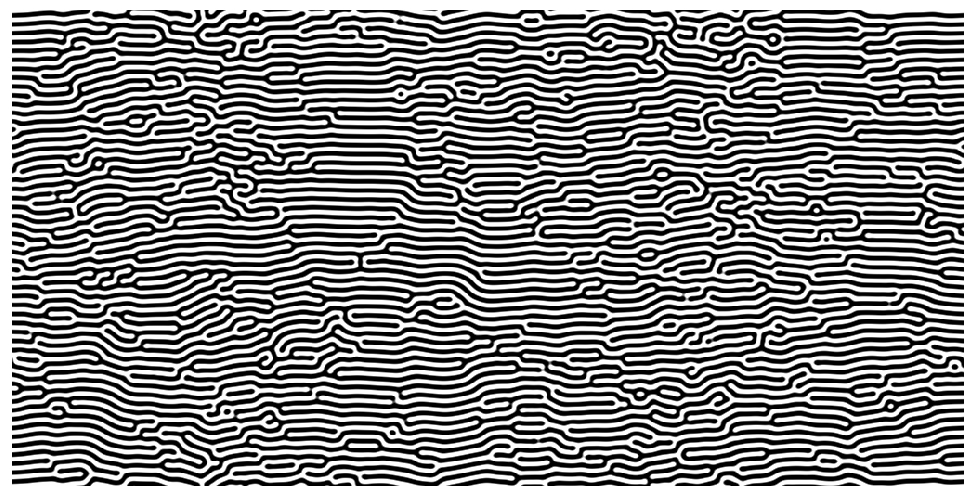}
\caption{Order parameter $\phi({\bf r})$ of system B-2D: after equilibration (timestep $t=3.2\e{5}$, top), in steady shear at $\dot{\gamma}=1\times10^{-4}$ ($t=8\e{5}$, centre) and after switch-off ($t=1.28\e{6}$, bottom).}
\label{fig2}
\end{figure}

Fig. \ref{fig2} shows equivalent states for system B-2D ($\beta = 0.42$).
The Peclet number during the shearing phase is now $Pe = 400$ $(Ca = 0.082)$.
All the generic results of Fig.~\ref{fig1} are the same, but the defect density, both during flow and after switchoff, is significantly reduced. (However the bands of misalignment in the flowing state are, if anything, even more apparent.) Consequently, the lamellar alignment along the flow direction after switch off is also more complete.

It is notable that at the shear rate shown in these figures, the more ordered system is B-2D which has higher $Pe$ than A-2D but smaller $Ca$. Varying $\dot\gamma$ within system A-2D or B-2D establishes unambiguously that the steady state order decreases with shear rate at fixed thermodynamic parameters (Fig.\ref{fig6} below). This points strongly to $Ca$ as the more relevant measure of flow rate when discussing onset of the disordered phase. 

In principle the dependence of the state of order on flow rate could be smooth, or could entail a nonequilibrium phase transition -- for instance if the perfectly ordered state ceased to be stable above a certain shear rate \cite{Gonnella98,Zilman99,Kumaran2001,Ramaswamy92}. In practice, however, starting as we do from an initially amorphous lamellar texture, a perfectly ordered state is generally not reached for large systems even if it might be stable once formed. In any case it seems likely that at high $Pe$, any defects introduced into the smectic state by deformation of an initially imperfect structure (whose layers are then dilated and compressed by the flow) cannot anneal away or annihilate each other rapidly. Such irregularities then build up until the steady state defect density is determined by a balance between their rate of introduction and rate of annealing. 

Recently Kumaran and Raman studied a single shear rate and varied the Schmidt number $Sc$ and system size, reporting enhanced alignment on decreasing $Sc$ and in small systems \cite{Kumaran2011}. Our comparative studies on systems A and C have confirmed both trends (although, as mentioned previously, our $Sc$ values are much higher). The same authors also discuss a distinct mechanism of defect creation caused by layer compression and splitting. Although we have not checked this in detail, this may be consistent with our observation of a strongly compressed and heterogeneous morphology, with bands oriented normal to the compression axis, as shown in Figs.~\ref{fig1} and \ref{fig2}.

A somewhat different morphology in sheared lamellar systems was reported by Xu et al.\cite{Xu06b}. This morphology exhibits macroscopic inhomogeneity between the centre of the simulated domain and the regions near the walls, accompanied by variations in the local shear rate. This contrasts with our own work which shows translationally invariant statistics in all directions. These differences may be a result of Xu et al.\cite{Xu06b} using mixed boundary conditions, as discussed already in the introduction, in which hard walls for the fluid flow are combined with periodic BCs on the order parameter field. 

\begin{figure}[htp]
\centering
\includegraphics[angle=0,width=0.5\textwidth]{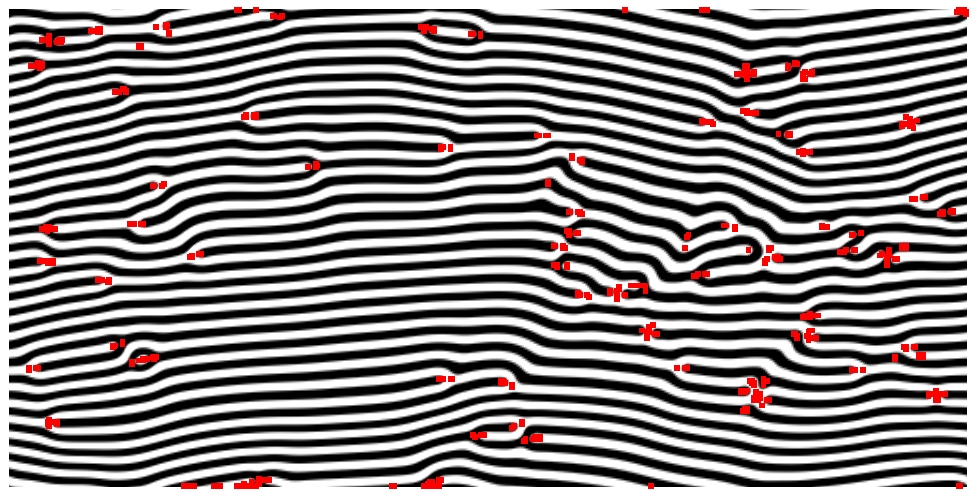}
\includegraphics[angle=0,width=0.5\textwidth]{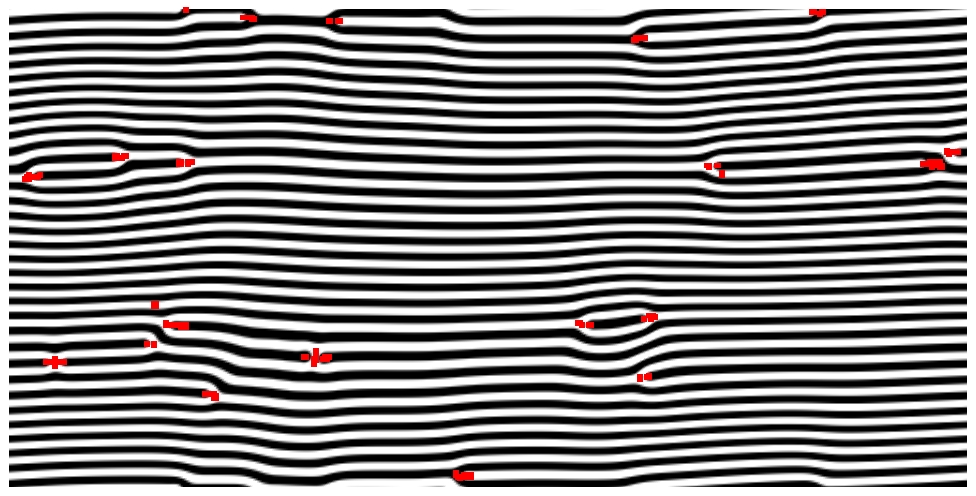}
\caption{Defect structure of system A-2D: at timestep $t=5.12\e{6}$ (top) and $1.024\e{7}$ (bottom) with an imposed shear rate $\dot{\gamma}=1.25\times10^{-5}$. The system has been pre-sheared at stepwise decreasing rates (see Table \ref{tab1}). Sites where the alignment function $X$ is smaller than $0.8$ are highlighted red. The system size was $N_x\times N_y \times N_z=512\times 256 \times 1$.}
\label{fig3}
\end{figure}

To quantify our observed morphology more closely, we use the alignment function $X$ introduced in Eq.~\ref{alignment}.
Fig.~\ref{fig3} shows snapshots of the time evolution of system A-2D at modest shear rate.
All lattice sites with $X({\mathbf r})\le0.8$ are marked in red and they obviously coincide with lamellar endpoints (edge dislocations) or other local defects.
The upper panel gives the situation at the lowest applied shear rate $\dot{\gamma}=1.25\e{-5}$, immediately following a pre-shearing protocol in which the strain rate was ramped down stepwise from higher values (see Table \ref{tab1}). While large aligned regions are present there are other regions of high defect density.
The lower panel of Fig.~\ref{fig3} shows a later state with a total strain incremented by a further $6400\%$ lattice units (i.e. by a displacement $64$ times the system size). 
Obviously many defects have annealed away and the smectic layers are now rather well aligned with the flow direction.

Defect annihilation to this large extent is only found at low enough shear rates. A prerequisite seems to be that the defects can smoothly approach each other and then annihilate. This is prevented by the buckling of layers above a certain shear rate; the undulations so formed ultimately pinch off and break, resulting in the creation of new defects. The onset of defect proliferation arises at characteristic flow rates $\dot{\gamma}_c=1.25-2.5\e{-5}$LBU in system A-2D and around $\dot{\gamma}_c=1-2\e{-4}$LBU in B-2D. 
(Here the subscript $c$ could stand for ``critical" or for ``crossover" depending on whether there is a singular transition or not; see above.)
These values are for
systems of size $512\times256\times1$; we found evidence of an increasing trend with system size, but the extremely long run-times involved ($\sim 10^7$ LB timesteps) precluded quantitative exploration. These results correspond respectively to values of the Peclet number, $Pe_c$, obeying $5<Pe_c<10$ (system A) and $400<Pe_c<800$ (system B).  However, in terms of $Ca = Pe/Pe^*$ one has $0.056<Ca_c<0.112$ (system A) and $0.082<Ca_c<0.164$ (system B). Given the 80-fold difference in $Pe_c$, the similarity of $Ca_c$ in the two different systems is remarkable.

Boldly extrapolating across parameter regimes that we have not explored in any detail, our results motivate the hypothesis that the onset of defect proliferation occurs generically in 2D at $Ca_c\simeq 0.08$.  Note that $Ca$, which comprises the ratio of a viscous stress to an elastic modulus, can also be viewed as a strain. In that context one can imagine that defects proliferate when the steady-state elastic deformation of the smectic exceeds a yield strain value (for which a threshold of a few percent is reasonable). This makes sense if, at $Ca \simeq Ca_c$, the stress contributed from the order parameter is (a) comparable to the viscous stress from the background fluid -- essentially the force balance argument made above -- and (b) primarily elastic in character. The latter in turn requires that $Pe_c$ is large and hence that $Pe^*$ is larger still. A different mechanism might apply at small $Pe^*$, since in that case layers relax rapidly by diffusion on the time scale set by $Ca_c$.

\begin{figure}[htp]
\centering
\hspace*{-0.1cm}\includegraphics[angle=0,width=0.5\textwidth]{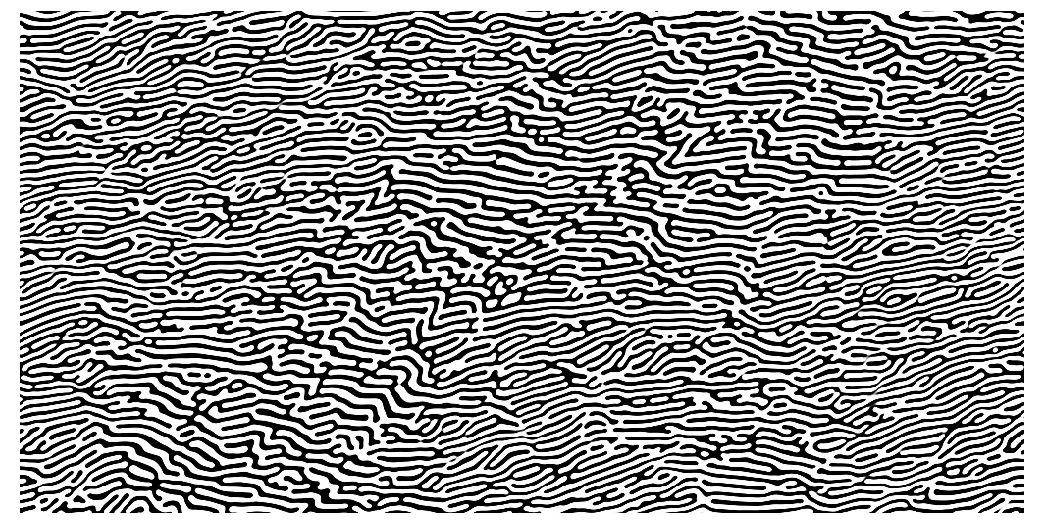}
\includegraphics[angle=0,width=0.49\textwidth]{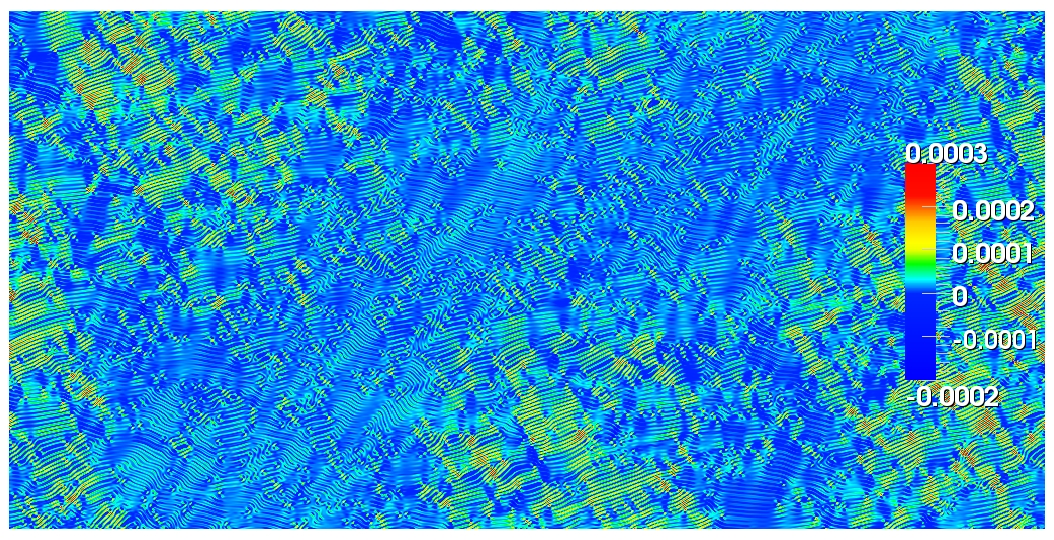}
\includegraphics[angle=0,width=0.49\textwidth]{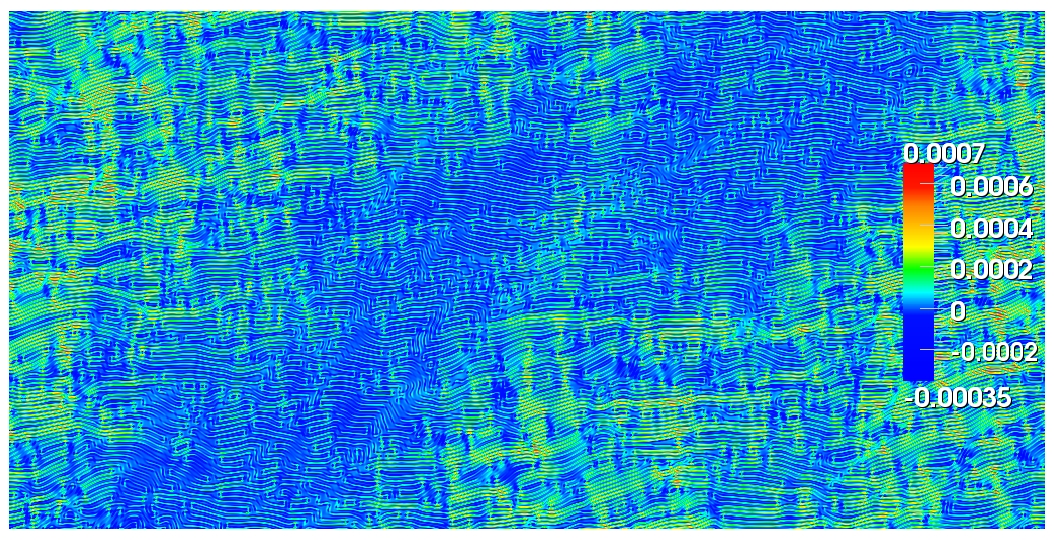}
\caption{Order parameter $\phi({\mathbf r})$ (top), shear stress $\sigma_{xy}({\mathbf r})$ (centre) and first normal stress differences $N_1({\mathbf r})$ (bottom) of system A-2D at $t=4\e{5}$ (see Table \ref{tab1}).}
\label{fig4}
\end{figure}

In Fig.~\ref{fig4} we explore the correlation between the inhomogeneous morphology in the flowing state and locally defined rheological quantities.
The pictures show the order parameter $\phi({\bm r})$, shear stress density $\sigma_{xy}({\bm r})$ and first normal stress difference density $N_1({\bm r})=\sigma_{xx}({\bm r})-\sigma_{yy}({\bm r})$ of system A-2D in steady shear flow at shear rate $\dot{\gamma}=1\e{-4}$ ($Pe = 40$, $Ca = 0.22$). Here and below we adopt a coordinate system where $x$ is along the flow direction and $y$ the velocity gradient.
In regions where the lamellar width is larger than in the equilibrium state, shear stresses and first normal stress differences show low positive or negative values.
Large positive values of both quantities coincide with regions where the lamellae are under compression.
Peak values (in LBU) are in the region of $-2\e{-4}$ and $3\e{-4}$ for shear stress and  $-3.5\e{-4}$ and $7\e{-4}$ for the normal stress differences.
Despite having locally negative regions, spatial and ensemble averages of these quantities are always positive.
Very similar results, but with smaller peak magnitudes, are obtained for system B-2D at the same shear rate.

\begin{figure}[htp]
\centering
\includegraphics[angle=0,width=0.33\textwidth]{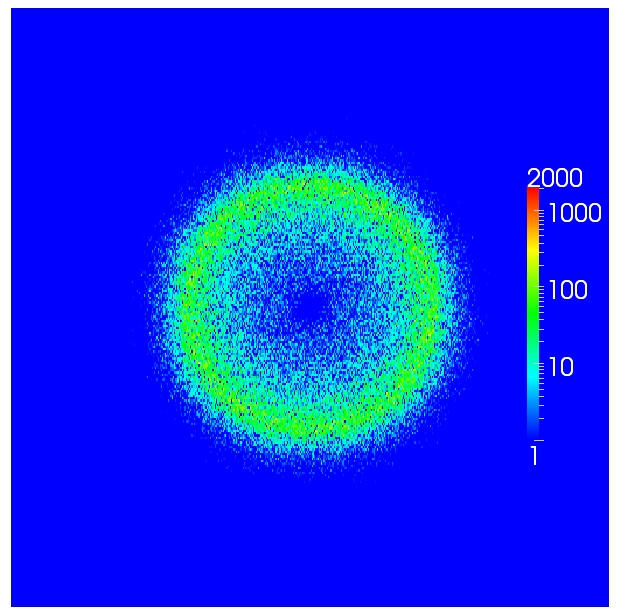}
\includegraphics[angle=0,width=0.33\textwidth]{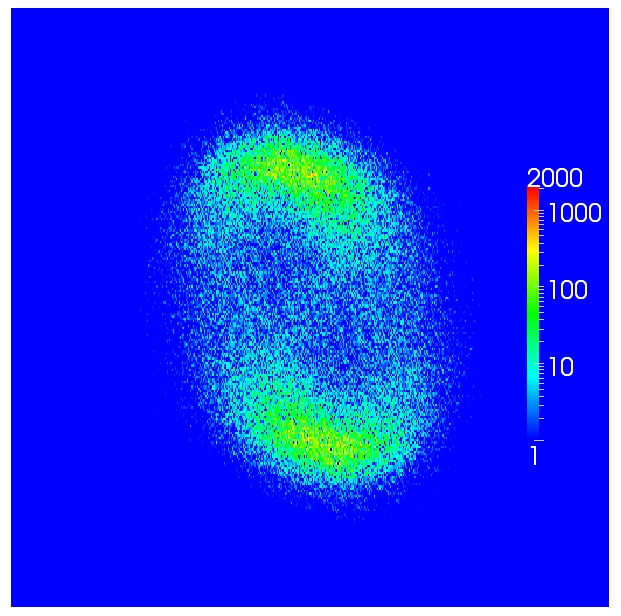}
\includegraphics[angle=0,width=0.33\textwidth]{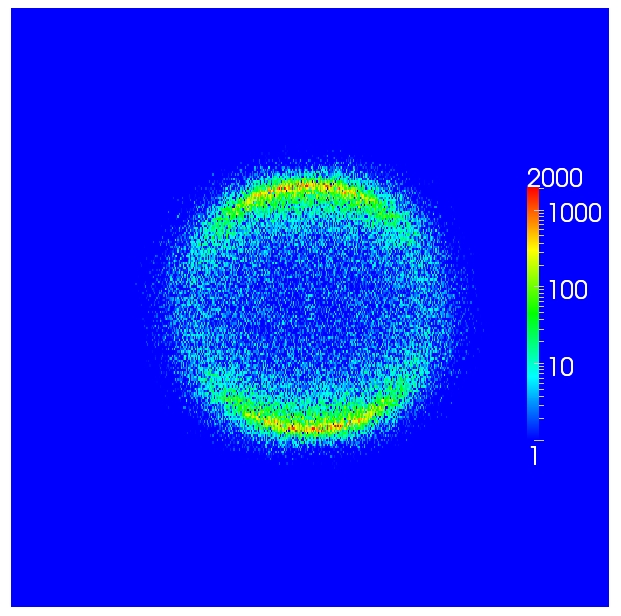}
\caption{Structure factor $C({\bm k})$ of system A-2D: Shown are states after equilibration at $t=3.2\e{5}$ (top); in steady shear at $\dot{\gamma}=1\times10^{-4}$ and $t=5\e{5}$ (centre) and after switch-off at $t=1.28\e{6}$ (bottom). The sections shown correspond to wave vectors on the interval $k_x, k_y \in [-\pi/2,\pi/2 ]$ LBU with $k_x$ and $k_y$ along the horizontal and vertical coordinate direction, respectively.}
\label{fig5}
\end{figure}

Fig.~\ref{fig5} shows structure factors $C({\bm k})$ for the same states as Fig.~\ref{fig1}.
$C({\bm k})$ is initially isotropic after layer formation and equilibration in the system at rest with, as expected, a distinct peak at $|{\bm k}|\simeq \pm2\pi/10$ LBU (recall $L = 10$ LBU).
In steady shear flow the structure factor develops two distinct but rather broad peaks at $k_x\simeq\mp 0.012\pi, k_y\simeq\pm 0.22 \pi$.
Thus the inhomogenous morphology of the lamellar structure is clearly encoded in the structure factor. 
The wavevector at the peak has a small component in the $x$-direction, indicating the average tilt of the lamellae away from the flow direction towards the elongational axis of the flow (which in shear is at $\pi/4$ to the flow direction). 

When the shear flow is switched off, the system relaxes again to its equilibrium layer spacing and the small tilt of the $C({\bm k})$ signal decreases to an undetectable level: this is consistent with the fact that, in Figs.~\ref{fig1} and \ref{fig2} (bottom panels) it is easy to see when the system has been sheared horizontally, but not in which sense. The intensity of the peak is slightly enhanced in the final state due to the improved orientation of the layers along the flow direction.

Fig.~\ref{fig6} shows the time evolution of the defects density $\rho_D$ of system A-2D in the sequence of stepwise decreasing shear rates alluded to previously (starting at $\gd=2\e{-4}$ and ranging down to $\gd=2.5\e{-5}$, Table \ref{tab1}).
The runs with the two highest shear rates were started from the amorphous, post-equilibration configuration, whereas all others were pre-sheared using the final configuration of the preceding run as the initial state.
The total strain between the beginning and end of every individual run was at least $3\times 10^{3}\%$.
Interestingly, following each step down in shear rate, both systems A-2D and B-2D continuously anneal to lower defect densities on a timescale $\simeq 10^6$LBU. If the shear rate is turned up again (not shown), the lamellae buckle and new defects recur on a timescale corresponding to a stain of about 100\%. 
This finding offers further support for a dynamical equilibrium between defect creation and annihilation above a critical (or crossover) shear rate, below which defects progressively annihiliate and near-perfect alignment eventually results.
To test this assumption we performed simulations starting from a very well aligned state and increased the shear rate.
The defect density indeed goes up (not shown) and reaches similar values to before. Note however that the defect annealing time (of order $10^6$LBU) is quite long compared to any inherent relaxation time; also, the final state after switchoff strongly depends on the previous flow. Thus the rheology and microstructure of these materials is inherently history dependent.

\begin{figure}[htp]
\centering
\includegraphics[angle=0,width=0.5\textwidth]{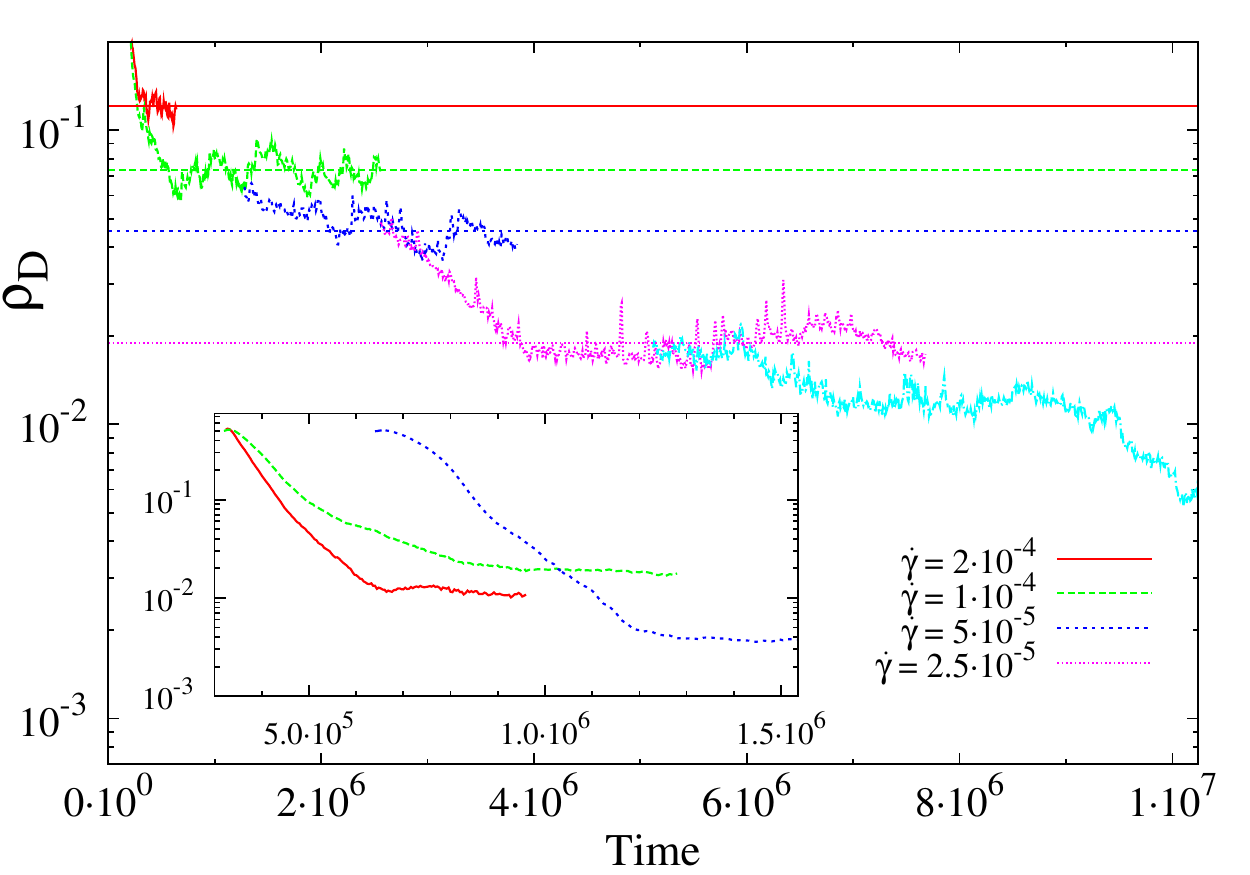}
\caption{Defect density $\rho_D$ vs time: The main picture shows data for system A-2D in a sequence of simulations with consecutively lowered shear rate $\dot{\gamma}=2\e{-4}$ (red solid), $1\e{-4}$ (green long-dashed),  $5\e{-5}$ (blue short-dashed),  $2.5\e{-5}$ (magenta dotted) and  $1.25\e{-5}$ (cyan dashed-dotted). Horizontal lines indicate time averages after entering the apparent steady state (where present). The inset gives data for the system A-3D and shear rates $\dot{\gamma}=2\times10^{-4}, 1\times10^{-4}$ and $5\times10^{-5}$ (left to right).} 
\label{fig6}
\end{figure}

\begin{figure}[htp!]
\centering
\includegraphics[angle=0,width=0.45\textwidth]{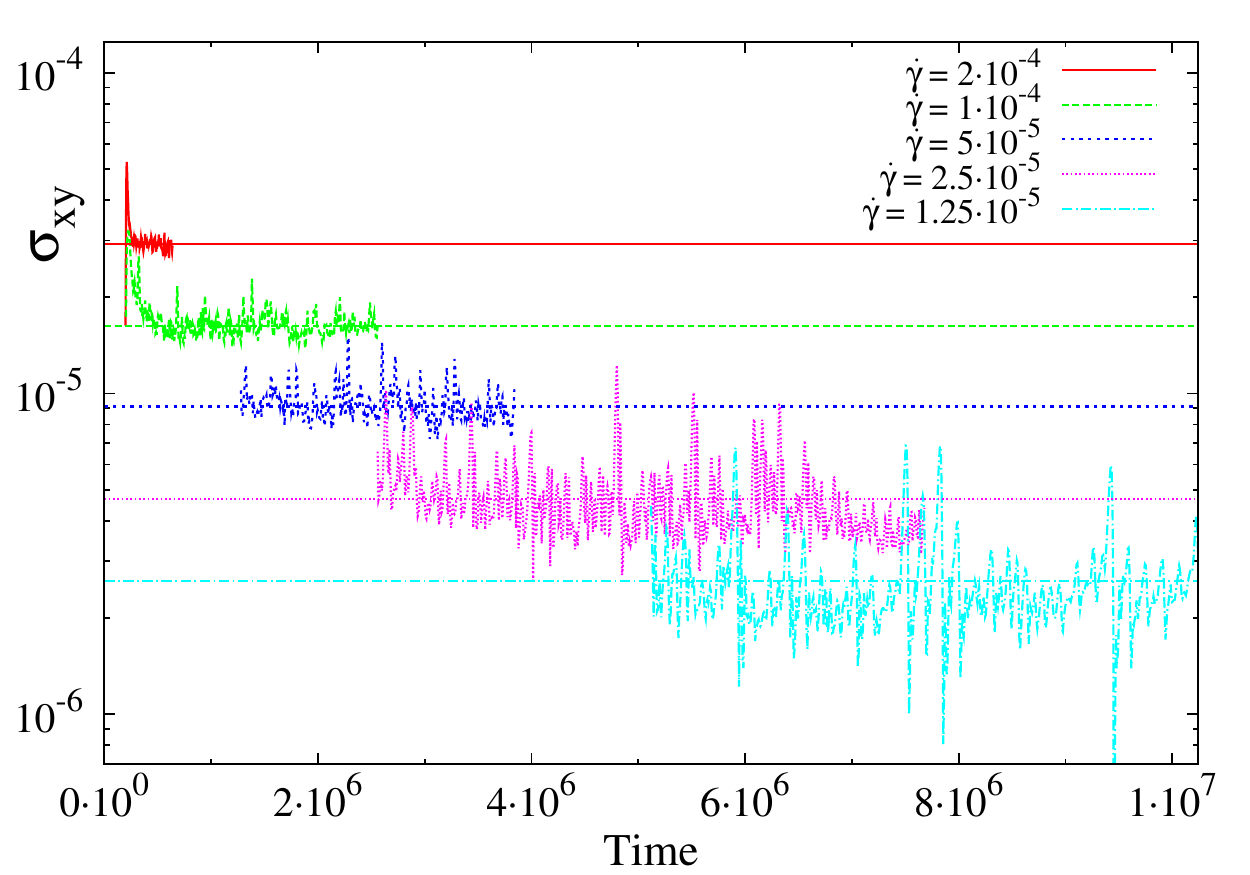}
\includegraphics[angle=0,width=0.45\textwidth]{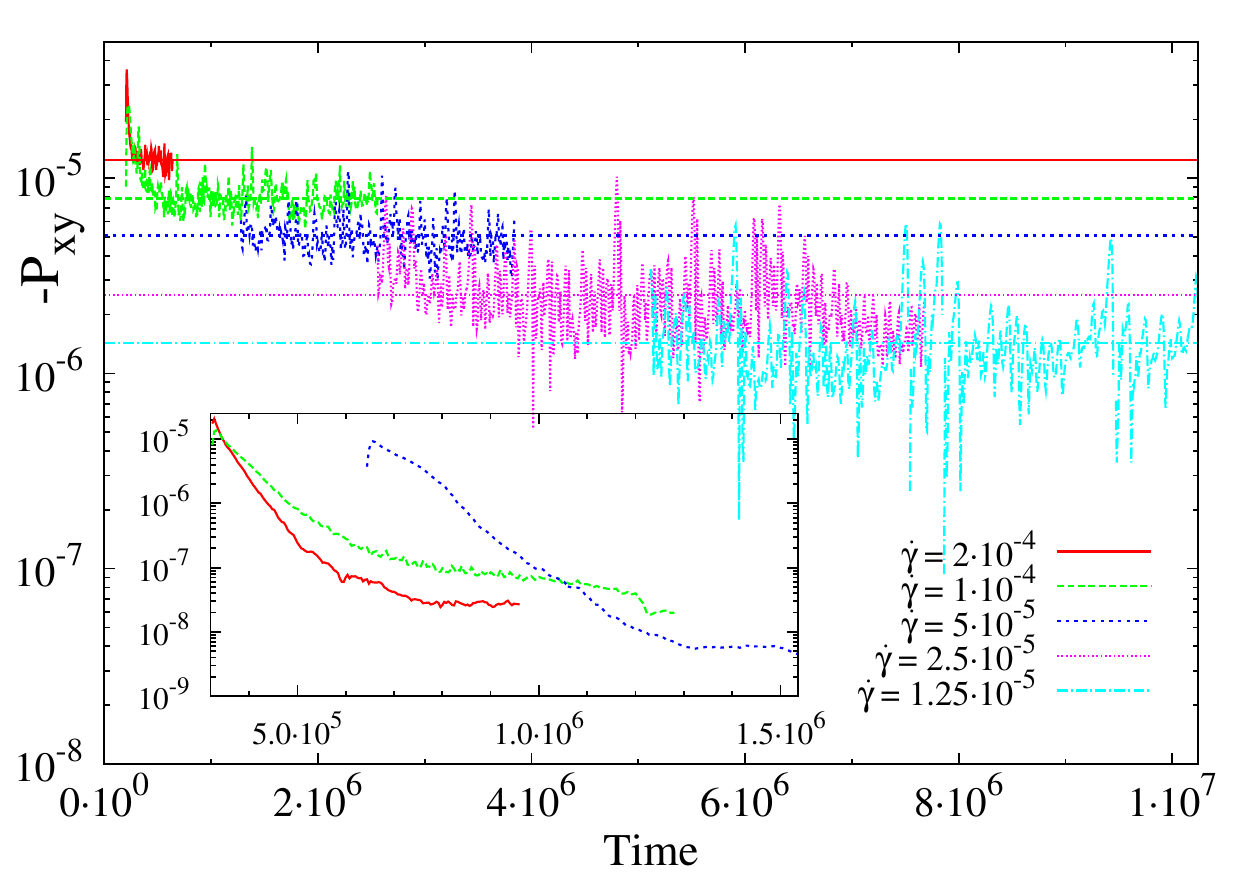}
\includegraphics[angle=0,width=0.465\textwidth]{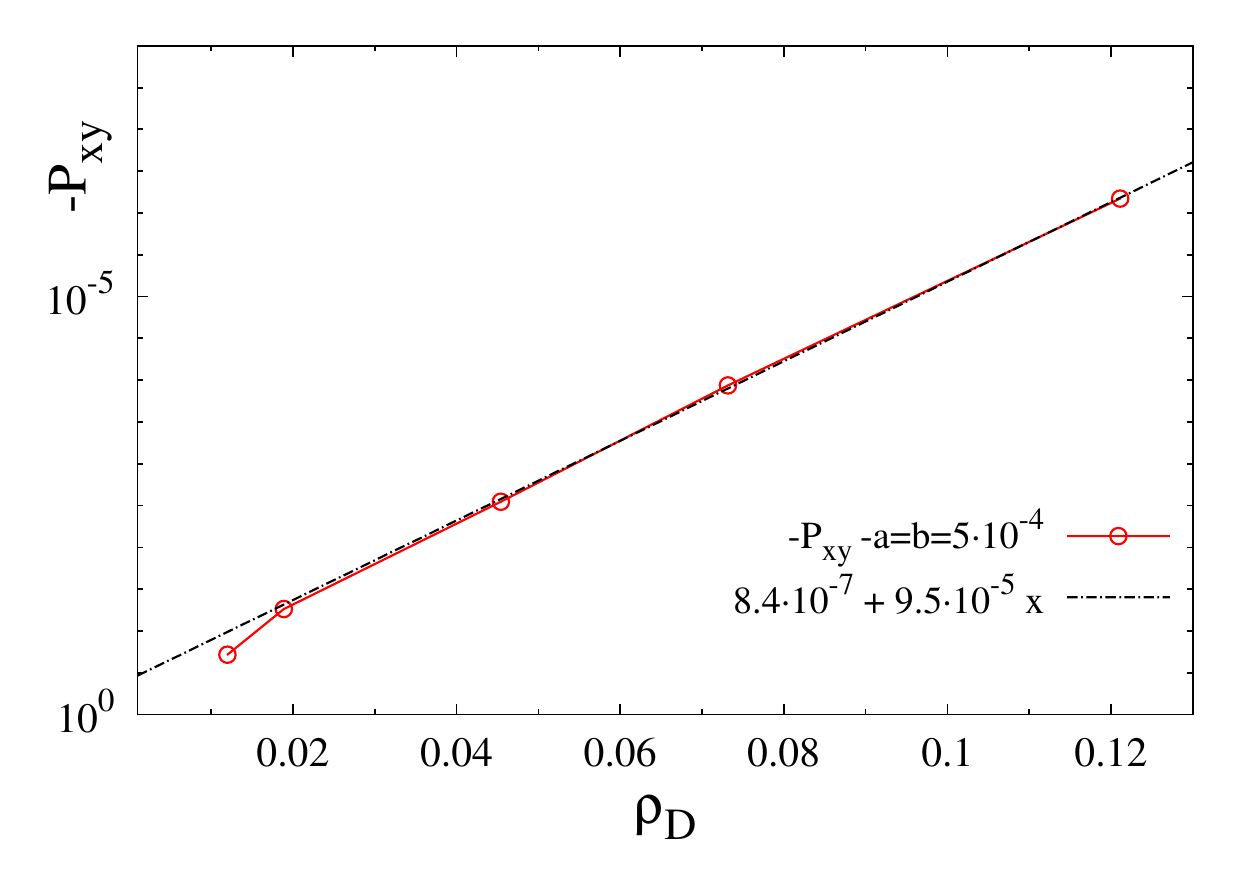}
\caption{Top: average total shear stress $\sigma_{xy}$ versus time for system A-2D: The horizontal lines are steady state time averages. For shear rate $\dot{\gamma}=1.25\e{-5}, 2.5\e{-5}, 5\e{-5}, 1\e{-4}, 2\e{-4}$, we find $\eta_{app}/\eta\equiv\sigma_{xy}/ \eta \dot{\gamma}=2.49, 2.25, 2.18, 1.93, 1.74$; centre: average order parameter contribution to the stress tensor $-P_{xy}$ vs time for system A-2D. The inset shows data for system A-3D and shear rates $\dot{\gamma}=2\times10^{-4}, 1\times10^{-4}$ and $5\times10^{-5}$ (left to right); bottom: average chemical contribution to the pressure tensor $-P_{xy}$ vs. defect density for system A-2D.}
\label{fig7}
\end{figure}

The top picture in Fig.~\ref{fig7} shows the spatially averaged total shear stress $\langle \sigma_{xy}({\bf r})\rangle \equiv\sigma_{xy} $.
At low shear rates, stress fluctuations have longer temporal correlations but smaller amplitudes. (The latter dependence is obscured by the logarithmic vertical scale in this plot.) The root-mean-square temporal fluctuations can be reduced by increasing system size (by a factor 3 or so for the runs shown in Fig. \ref{fig1} and Fig. \ref{fig2}). In contrast, the averaged shear stress is almost independent of the system size, which makes it possible to simulate longer times using smaller system sizes while still giving accurate predictions for the bulk rheological behaviour. 
We pursue this strategy here.

System A-2D shows clear shear thinning behaviour; values of its apparent viscosity at various shear rates are given in the caption of Fig. 7. The corresponding range of $Ca$ is 0.028-0.448. Using Eq.\ref{fullstress}, we can identify the stress contribution arising from the deviatoric part of the order parameter pressure tensor
$\langle - P_{xy}({\bf r})\rangle\equiv - P_{xy}$; this is shown in the centre graph in Fig.~\ref{fig7}. The temporal fluctuations are in phase with those of the total stress and decrease with decreasing shear rate.
The bottom image in Fig. \ref{fig7} plots $- P_{xy}$ against the defect density $\rho_D$; this shows a linear relation across almost the whole range of shear rates.

In contrast, system B-2D does not exhibit clear shear thinning; the apparent viscosities obey $\eta_{app}/\eta=1.57, 1.62, 1.55, 1.42, 1.87$ for the same set of shear rates as used in Fig.~\ref{fig7}. To within the measured accuracy, system B-2D is thus rheologically quasi-Newtonian. However, the range of $Ca$ in this case is lower, 0.01-0.164, and the order parameter stress creates only a relatively modest uplift to the baseline viscosity. More significantly, $Ca\le Ca_c$ for all but the highest applied shear rate so that the steady states found in system B-2D are far more ordered than in A-2D. This results from a defect annihilation process, occurring in the early stages of ramp-down, which we found to be accompanied by a sharp drop in $P_{xy}$ (not shown).

\begin{figure}[htp!]
\centering
\includegraphics[angle=0,width=0.5\textwidth]{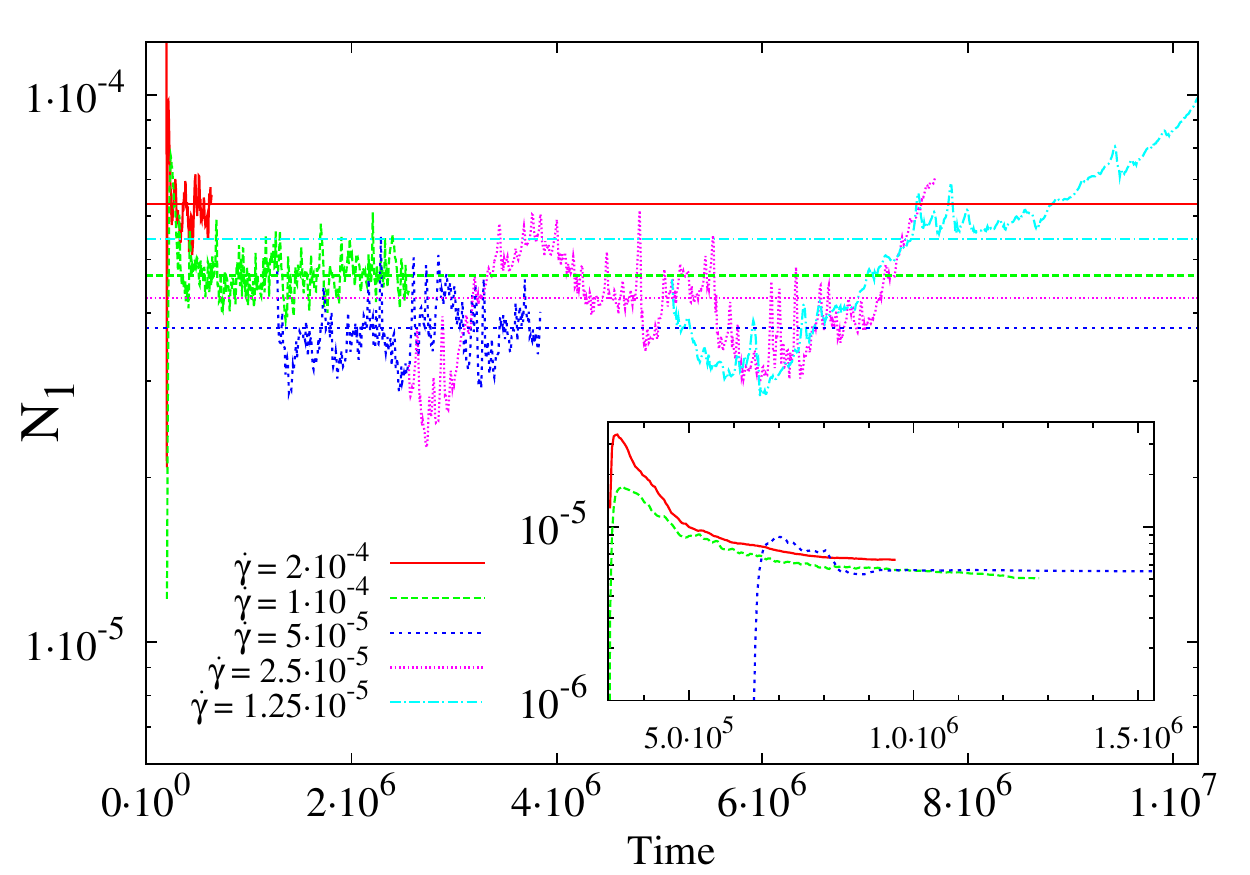}
\caption{Spatially averaged first normal stress differences $N_1= \sigma_{xx}-\sigma_{yy}$ vs time for system A-2D. The inset shows data for system A-3D and shear rates $\dot{\gamma}=2\times10^{-4}, 1\times10^{-4}$ and $5\times10^{-5}$ (left to right). Horizontal solid lines are time averages.}
\label{fig8}
\end{figure}

Fig.~\ref{fig8} shows averaged first normal stress differences $\langle N_1({\bf r})\rangle \equiv N_1$ versus time for system A-2D, where we report the same stepwise decreasing flow protocol as before.
For the first several steps $N_1$ decreases monotonously with decreasing shear rate.
However, at the lowest shear rates the behaviour is not monotic: $N_1$ eventually rises again. (The resulting time averages, shown as horizontal lines, do not therefore reflect true steady states.)
This somewhat surprising feature has a simple explanation. 
The time sequence of the defect density in Fig. \ref{fig6} reveals that the rise in $N_1$ is correlated with a drop in $\rho_D$, and linked to the onset of defect annihilation (which happens only at low shear rates).
When two defects meet and cancel, the local lamellar width alters by a certain amount. This causes frustration as the system is unable to snap back locally towards its equilibrium layer spacing. Hence the normal stress difference $N_1=\sigma_{xx}-\sigma_{yy}$ fails to decay to zero, and remains positive at large times, as appropriate for a system of metastable and (on average) compressed smectic layers. The time dependence of the normal stress difference is also anomalously slow, whereas the shear stress reaches their steady state values more quickly. The relatively large steady state value of $N_1$ and this slow dynamics may partly be a finite size effect caused by the relatively modest number of smectic layers within the simulated volume.
The normal stress in system B-2D (not shown) behaves differently: in this case $N_1$ falls monotonically during the defect annihilation phase. 
It is not clear whether this difference should be attributed to the change in $\beta, Ca$ or $Pe$ between the two systems, or indeed a combination of those changes.

\begin{figure}[htp!]
\centering
\includegraphics[angle=0,width=0.5\textwidth]{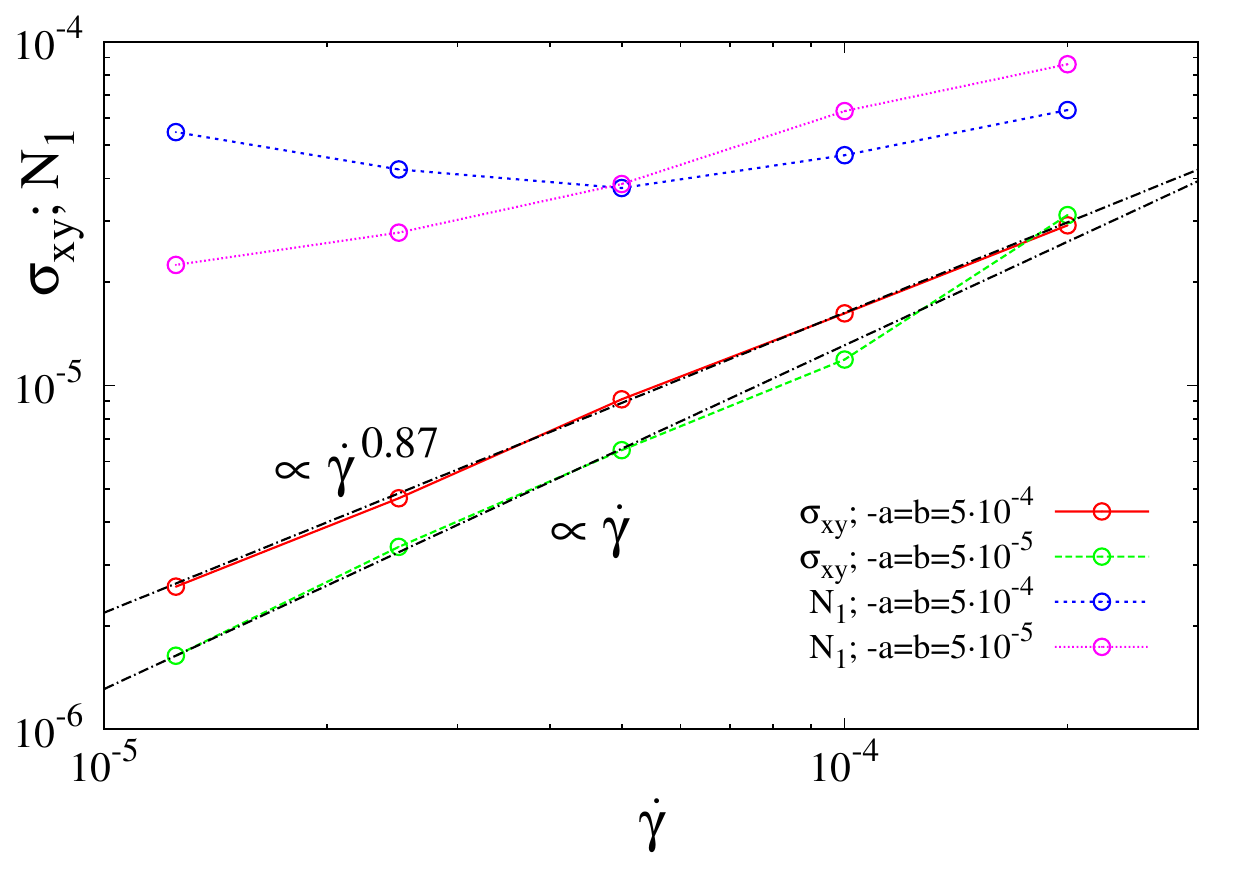}
\caption{Flow curves: Time-averaged shear stress $\langle \sigma_{xy}\rangle$ and first normal stress differences $\langle N_1 \rangle$ vs shear rate $\dot{\gamma}$ for system A-2D and B-2D. The dash-dotted lines are power-law fits to the data.} 
\label{fig9}
\end{figure}

Fig.~\ref{fig9} shows `flow curves' representing the shear rate dependence of time averaged stresses $\sigma_{xy}$ and $N_1$. This is calculated from the stepwise ramp data as detailed previously, so the $N_1$ results at low shear rates should not be taken as definitive. System A-2D shows shear thinning as already discussed, apparently following a power law $\sigma_{xy}=\dot{\gamma}^m$ with exponent $m\simeq 0.87$. A recent experimental and theoretical work studies the nonlinear relation between shear stress and shear rate in lyotropic lamellar phases \cite{Lu08}; interestingly the authors identify the motion of screw defects as being responsible for the layer tilting, folding  and formation of defects. They report an experimental value of $m\simeq 1/1.44\simeq 0.694$, which is in good agreement with their theoretical prediction of $m=2/3$. Our value of $m=0.87$ for the disordered flow regime is larger. One possible explanation for this discrepancy is that screw defects, which do not exist in our 2D simulations, play a key role for this theory. 
On the other hand our fitted power law must be treated with great caution since the fit includes data from $\dot\gamma$ both above and below $\dot\gamma_c$. It is somewhat surprising that the morphological changes around $1.25\e{-5}<\dot\gamma_c <2.5\e{-5}$ in system A give no visible anomaly or feature in the flow curves.
%
For system B-2D the data is consistent with an exponent $m=1$, indicating more or less constant viscosity and no pronounced shear thinning (as previously described). 
This appears to be due to the decreasing number of defects at shear rates below $\dot{\gamma}\le2\e{-4}$; the order parameter stress $P_{xy}$ is then dominated by the background viscosity term $\eta\dot\gamma$. 

{In neither system does $\sigma_{xy}(\dot\gamma)$ offer any clear signature of a yield stress. There may be two ways to understand this. Either, as detailed
previously, this is because at the lowest shear rates studied here the selected
state is a fully aligned one with the layering wavevector perpendicular to
the flow: such a state should have a purely viscous response. However, a defect texture may also lead to a viscous response in the linear regime if flow occurs via permeation~\cite{deGennes} (i.e. by fluid flow not altering appreciably the order parameter pattern). 

To further characterise the small shear rate regimes in our simulations, we have performed additional simulations on fully aligned states, with different layer spacings. While the shear stress contribution in this case is negligible as $\dot{\gamma}\to 0$, $N_1$ does not vanish for vanishing shear rate, but depends sensitively on the nonequilibrium layer spacing in the range of shear rates addressed here.  This can be described as a ``quasi-Newtonian'' behaviour. As expected, $N_1$ is positive for overcompressed layers, and negative for overdilated layers. The values of $N_1$ obtained in system B-2D (see Fig.~9) are comparable to those observed in a quiescent lamellar smectic system prepared by compressing the layers by about 5-10\%.}

\subsection{Three Dimensions}
\label{results3D}

The computational cost of repeating all the above studies in three dimensional systems of equivalent size remains prohibitive, particularly when addressing steady state properties which emerge only after a million or more LB timesteps. We therefore focus in 3D on smaller systems (Table \ref{tab1}), but still large enough in principle to show qualitatively similar structure under flow. Our studies use two sets of parameters. System A-3D is identical to A-2D except for the introduction of the third dimension. System C-3D has parameters intermediate between A-2D and B-2D (Table \ref{tab1}) and was chosen in preference to the latter because B-2D ordered completely if the system size was matched to a slice of B-3D. (We see below however that in fact this is a rather general outcome in 3D.)

{
\begin{figure*}[htp!]
\centering
\includegraphics[angle=0,width=0.45\textwidth]{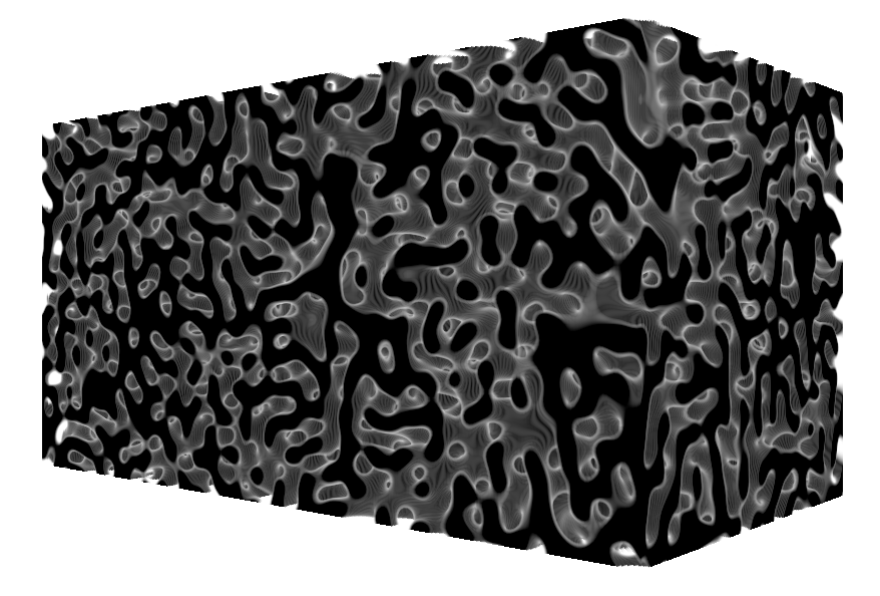}
\includegraphics[angle=0,width=0.45\textwidth]{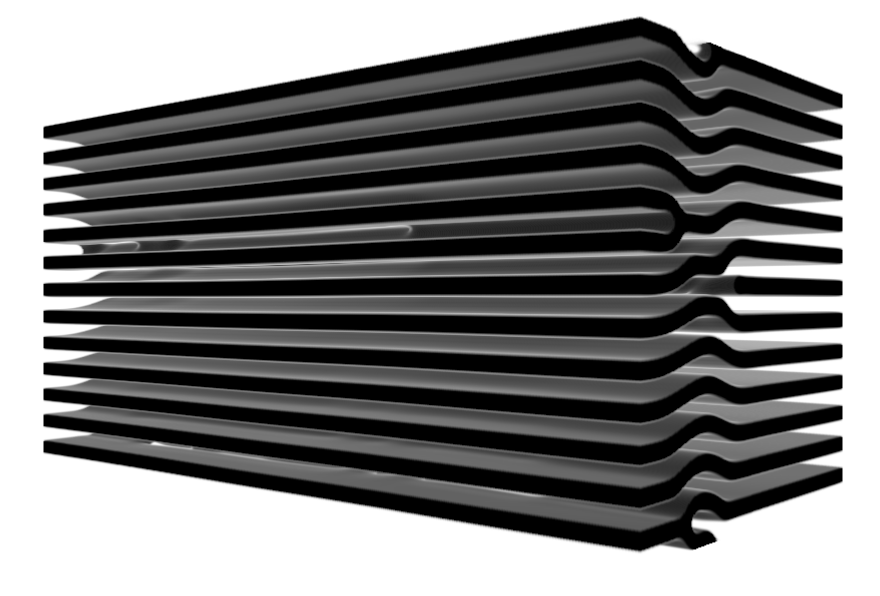}
\includegraphics[angle=0,width=0.45\textwidth]{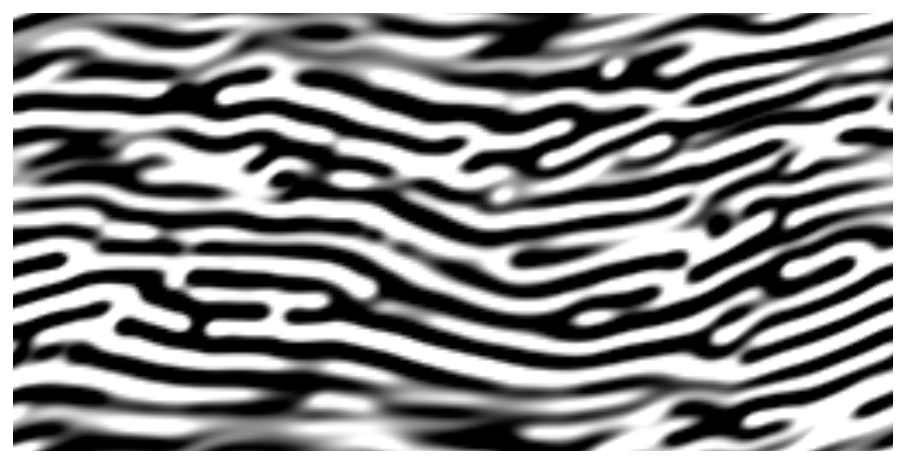}
\includegraphics[angle=0,width=0.45\textwidth]{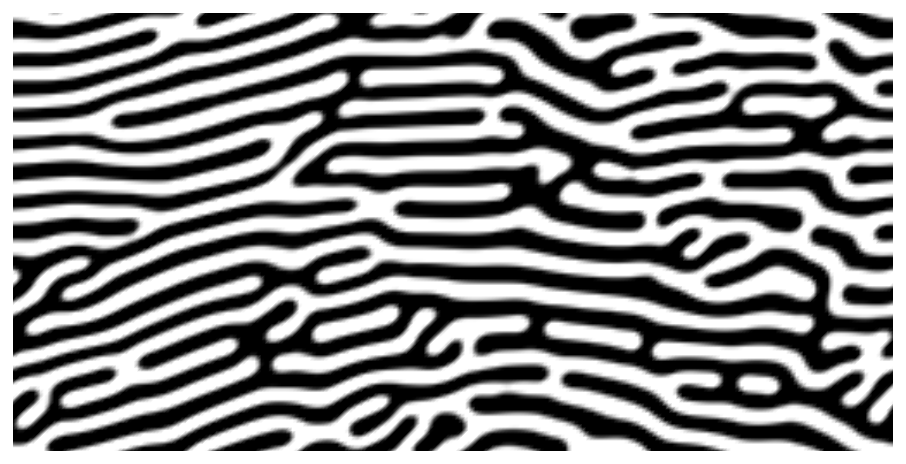}
\caption{Order parameter $\phi({\bf r})$ in two and three dimensions: Top pictures show system C-3D after equilibration (left) and in shear flow (right, $\dot{\gamma}=2\times10^{-4}$) at $t=6.4\e{5}$ LBU. The opacity of one phase has been reduced to enable a better view into the lamellar structure. The bottom left picture shows system C-2D at the same time step and shear rate. The bottom right picture is system A-2D at $\dot{\gamma}=1\times10^{-4}$ and time step $t=9.6\e{5}$ LBU. Results for A-3D are morphologically indistinguishable from those for C-3D. In all these images the flow direction is horizontal, to the right in the upper half of the sample and to the left in the lower half.} 
\label{fig10}
\end{figure*}
}

\begin{figure}[htp!]
\centering
\includegraphics[angle=0,width=0.40\textwidth]{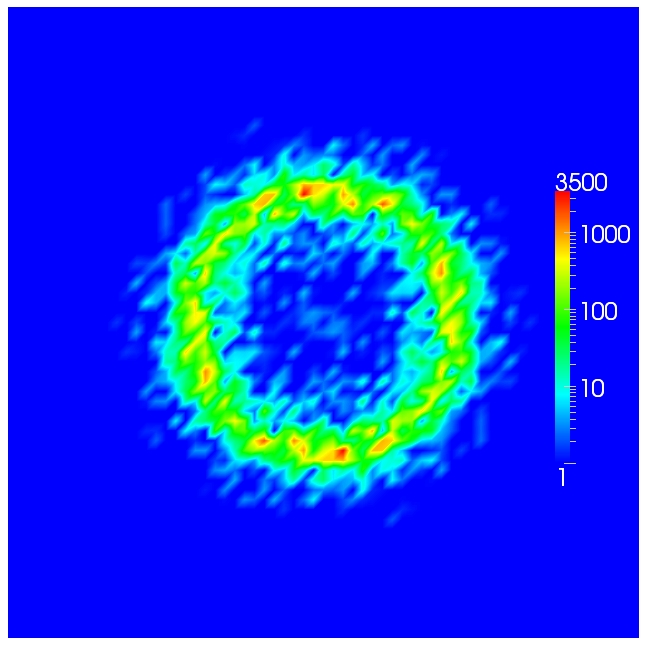}
\caption{Structure factor $C({\mathbf k})$ of quiescent system A-3D, showing a cut along $k_x=0$ after equilibration at $t=3.2\e{5}$. Shown are wave vectors on the interval $k_y, k_z\in[-\pi/2 ,\pi/2 ]$ with $k_y$ and $k_z$ being the horizontal and vertical direction, respectively. Data for C-3D (not shown) is virtually the same.}
\label{fig11}
\end{figure}

\begin{figure}[htp!]
\centering
\includegraphics[angle=0,width=0.35\textwidth]{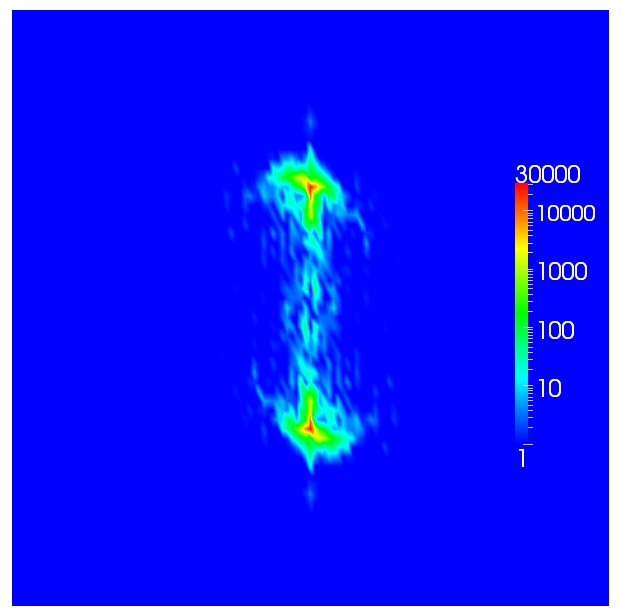}
\includegraphics[angle=0,width=0.35\textwidth]{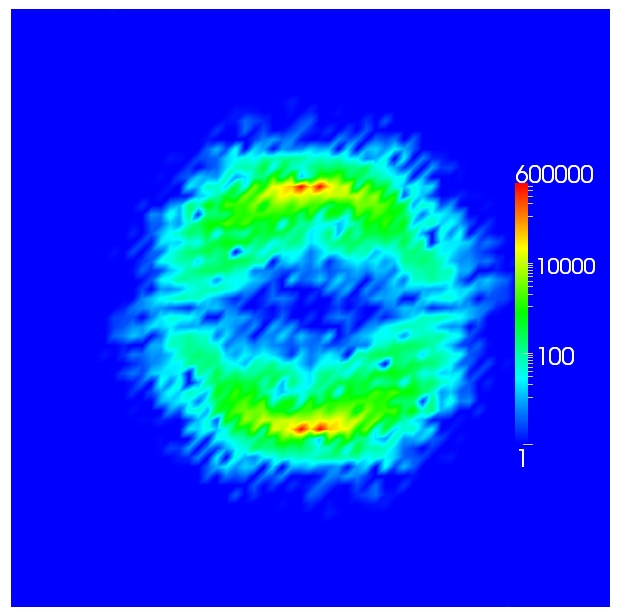}
\includegraphics[angle=0,width=0.35\textwidth]{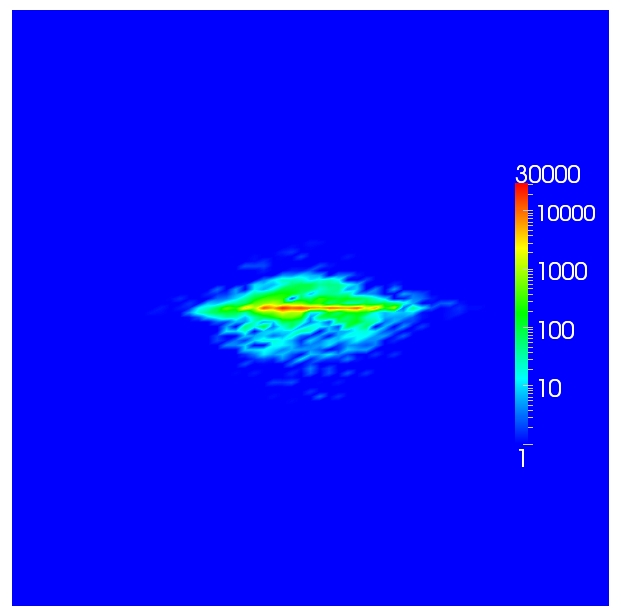}
\caption{Structure factor $C({\mathbf k})$ of system A-3D in shear flow ($\dot{\gamma}=1\times10^{-4}$) at t=960k. Results are shown in the planes at $k_x=0$ (top panel), $k_z=0$ (middle panel), and $k_y=3\pi/16$ (middle panel). 
The horizontal and vertical directions are respectively along $k_z$ and $k_y$ (top panel), $k_x$ and $k_y$ (middle panel), $k_x$ and $k_z$ (bottom panel). The wave-vectors in the horizontal and vertical directions are in $[-\pi/2 ,\pi/2]$ for all panels. The structure factor is akin to that expected of a regular lamellar phase.}
\label{fig12}
\end{figure}

Fig.~\ref{fig10} shows the post-equilibration amorphous structure and the steady-state structure under flow in system C-3D (A-3D is similar) at system size $N_x\times N_y \times N_z=256\times128\times128$. The latter is compared with C-2D and A-2D in a system of the same $x$ and $y$ dimensions ($256\times128\times1$).  The quiescent metastable state of system C-3D after equilibration is more or less the same as it would be in two dimensions. 
Indeed, a slice through the structure factor along $k_x=0$ (cuts along $k_y,k_z=0$ look identical) is shown in Fig. \ref{fig11} and resembles that in
Fig.~\ref{fig5}.
However, under shear at $\dot\gamma = 2\times 10^{-4}$ ($Pe = 400$, $Ca = 0.252$) this structure anneals readily into a regular lamellar structure after a transient period of about $6\e{5}$ LB timesteps. This occurs for even the highest shear rates applied, indicating that there is a very strong tendency in three dimensions to form ordered lamellar arrays. The same behaviour was seen in system A-3D for which the highest flow rate used corresponds to $Pe = 80$ and $Ca = 0.448$. (In contrast, the 2D results at the same flow rates and system sizes feature the typical fluctuating pattern of breaking and merging lamellae that was described previously and is visible in the lower panels of Fig.~\ref{fig10}.) The structure factor seen in 3D (see Fig.~\ref{fig12}) is accordingly close to that of a perfectly ordered lamellar state, in contrast to the corresponding results in 2D (Fig.~\ref{fig5}).
These findings suggest that any characteristic strain rate $\dot\gamma_c$ at which defects proliferate must be markedly higher in 3D than in 2D. Specifically, if $Ca$ is the relevant control parameter, our results for A-3D imply that $Ca_c > 0.448 $ in three dimensions.

Note however that the order we observe is not completely perfect as a mismatch of the layers occurs leading to branching points and edges (e.g. at the centre of the ordered 3D-system in Fig.~\ref{fig10}). 
The Burgers vector is oriented along the gradient direction with edges along the flow direction.
This forms a structure that is translationally invariant along the flow direction and in the final steady state the location of the edge defects does not change. Notably, the corresponding orientation of edge dislocations does not exist in two dimensions.

{A closer look at the kinetic evolution of the order parameter reveals that the additional dimension appears to play a key role in promoting the formation of the homogeneously layered structure.
The ordering process takes place in two consecutive steps.
Firstly, the equilibrated state undergoes a transition to a semi-ordered state, which consists of small patches of layered regions, with the layer normals along gradient direction.
The length of the lamellae is initially quite short, but gets longer along the flow direction as the simulation proceeds and the structure is advected. It remains, however, relatively short along the vorticity direction, with the individual layers being disconnected.
In a second step the layers merge by linking up along the vorticity direction. This is possible as the layers can be locally translated along the flow gradient direction, so that the average order parameter within a plane can change and is not conserved as in the two-dimensional case. }

We next quantify the time system A-3D takes to anneal into its final ordered structure starting from an isotropic initial configuration. (These and subsequent measurements are reported for A-3D only; C-3D is qualitatively similar.)  
The inset of Fig.~\ref{fig6} gives the defect density as defined by Eq.~\ref{defect density} versus time for shear rates $\dot{\gamma}=5\times10^{-5}-2\times10^{-4}$. 
Within a few hundred thousand LB timesteps the defect density decreases by two orders of magnitude, resembling the behaviour seen in system A-2D at shear rates one decade smaller.
The total strain where the curves level off, and ordering is achieved is about $\gamma=160$ (at $\dot{\gamma}=2\times10^{-4}$), $\gamma=120$ (at $\dot{\gamma}=1\times10^{-4}$) and $\gamma=65$ (at $\dot{\gamma}=5\times10^{-5}$).

The reduced number of defects is reflected in the magnitude of the order parameter stress contribution $P_{xy}$; the 3D behaviour is shown in the inset of Fig.~\ref{fig8}. The final contribution to the total stress is created by a small number of remaining defects (hence the inconsistent trend of the tiny plateau value with strain rate, which is not a concern here) and is quite negligible compared to the background viscosity contribution $\eta\dot\gamma$. The drop in $P_{xy}$ by three orders of magnitude during the defect annihilation process can formally be viewed as shear thinning of a quite extreme degree -- albeit one masked, in any macroscopic measurement of $\sigma_{xy}$, by the Newtonian background. 

First normal stress differences $N_1=\langle\sigma_{xx}-\sigma_{yy}\rangle$ for system A-3D are shown in the inset of Fig.~\ref{fig8}.
For all shear rates addressed here system A-3D attains similar values of the normal stress; these are approached monotonically and about one decade smaller than those seen in system A-2D. The mechanism reported previously for the buildup of normal stress on approaching the ordered state in 2D systems seemingly does not apply in 3D, perhaps because of the much more efficient pathways to relaxation and mutual annihilation of defects.

\if{ 
Finally we want to glimpse at the structure factor of system A-3D in shear flow.
Fig. \ref{fig12} shows $C({\bm k})$ at shear rate $\gd=1\e{-4}$ for Fourier vectors on the interval $[-\pi/2 l, \pi/ 2 l]$.
The leftmost picture is a cut along $k_x=0$ in with $k_y$ in horizontal and $k_z$ in vertical direction.
It shows data in flow-velocity-gradient plane and can be directly compared with the left picture in Fig. \ref{fig5}.
The other images give cuts along $k_y=0$ in $k_x-k_z$-plane and along $k_z$ in $k_x-k_y$-plane. 
Note that for the centre picture a logarithmic colour map had to be adopted due to the strong ordering in 3D.
There is a certain similarity between the 2D-structure factor {\it after} switch-off (Fig. \ref{fig5}) and the 3D-structure, but the latter shows a pronounced peak at $k_z\simeq\pm0.2\pi/l$ also during shear flow, which is slightly concealed by the logarithmic scale.The split nature is an effect of the relatively sparse discretization and was caused by the visualisation software.
Although strongly peaked, the spatial dependence of the signal is best demonstrated in a 3D-representation. 
Fig. \ref{fig13} depicts a close up of the 3D-structure factor, whereas the red area corresponds to an isosurface of about $C({\bm k})\simeq 10^3$.
It reveals a banana-shaped spatial dependence with some small appendices in flow direction at $k_x=0$.

\begin{figure}[htp!]
\centering
\includegraphics[angle=0,width=0.45\textwidth]{ck_run788_960_zoom.jpg}
\caption{Structure factor $C({\mathbf k})$ of system A-3D in shear flow: The picture gives a 3D-representation of the data in shear flow at $t=9.6\e{5}$ and $\dot{\gamma}=1\times10^{-4}$.}
\label{fig13}
\end{figure}

}\fi

\section{Conclusions}\label{conclusions}

Many earlier studies have proposed a picture in which a layer undulation instability in three-dimensional smectic liquid crystals emerges above a critical shear rate \cite{Zilman99,Courbin02,Gonnella98,Kumaran2001,Ramaswamy92}. Molecular dynamics simulations on lyotropic smectics place this in a regime of flow rates well above those we have applied in the present study \cite{Guo2002,Soddemann2004}. The critical shear rate for thermotropic smectics that has been given in a theoretical approach \cite{Stewart2009} is slightly above ours. The status of this instability and its effects on rheology in experiments remains somewhat unclear. Also unclear is its status in two dimensions.

In the present work we have studied the nonlinear rheology of lamellar systems in two and three dimensions by means of large-scale simulations of a well-established model. After quenching from a homogeneous phase a metastable amorphous lamellar structure appears, similar to previous simulations in two dimensions using comparable models and parameters~\cite{Gonnella97,Gonnella98,Xu03,Xu05,Xu06a,Xu06b}.
Also in accord with some previous studies, under flow in two dimensions we found the morphology of the order parameter to show two distinct regimes: one with very few defects which arises at low shear rates, and a defect rich phase at higher ones. We found that the critical (or crossover) strain rate $\dot\gamma_c$ separating these phases depends somewhat on system size, but much more strongly on thermodynamic and kinetic parameters. We have argued that this dependence might be summarized in terms of a capillary number $Ca$ (Eq.~\ref{capillary}) that is the ratio of viscous stress to the gradient free energy density of the lamellar phase. This controls a steady-state elastic strain in the smectic which, when it exceeds some characteristic value $Ca_c$, might cause yielding of layers and defect proliferation; our estimate is $0.056<Ca_c<0.112$ (system A) and $0.082<Ca_c<0.164$ (system B). However we emphasize that the evidence for $Ca$ as a control parameter is currently very limited and it would take a painstaking exploration of parameter space to confirm this picture. 

A notable feature of the disordered flow phase is our observation of striated heterogeneity in the state of smectic order, with strongly dilated and compressed regions oriented along the extensional axis of the flow.  These are present in steady state but are nonstationary. The extended regions with lamellar spacing larger than at equilibrium exhibit either negative or small positive shear stress and first normal stress difference, whereas positive peak values were found in the compressed domains.
(A similar dependence of the layer alignment in flow direction depending on the system size has been recently reported in \cite{Kumaran2011}.)
Smaller systems with 16 layers across the gradient direction, which is comparable with our systems consisting of  $N_x\times N_y \times N_z=256\times128\times1$ sites, showed a larger tendency to align in flow direction.
For larger systems there was a clear trend to align at an angle with the flow direction.
These distortions of the order parameter contribute a term in $P_{xy}$ to the shear stress whose magnitude could be directly related to the density of defects. In one system we found shear thinning described by a power law dependence of stress on strain rate, $\sigma_{xy}=\gamma^m$ with $m=0.87$; however the simplicity of this result may be fortuitous since the range of strain rates studied spans $\dot\gamma_c$. In another system, at strain rates primarily below $\dot\gamma_c$, the shear rheology was quasi-Newtonian ($m=1$) although a significant first normal stress difference $N_1$ was also present.

We then simulated the flow behaviour of directly comparable systems in three dimensions, and found striking differences between the two- and three-dimensional response under conditions of matched $\dot\gamma$. The 3D-system showed none of the distictive dilation/compression bands seen in 2D; moreover the microstructure, initially amorphous after quenching from the homogeneous phase, rapidly ordered under shear into a lamellar array with a low density of edge defects, all of which were oriented along the flow directions. This contrasts with the 2D case where numerous edge dislocations are present in the disordered phase ($\dot\gamma > \dot\gamma_c$) but, by virtue of the 2D geometry, all are oriented along the vorticity direction. (In saying this, we imagine a 3D continuation of the 2D structure which is translationally invariant along that direction.) Our 3D results do not rule out a transition or crossover from ordered to disordered structures, but this is not seen within the range of flow rates addressed here. This means that, if such a transition is controlled by capillary number, its value at onset in 3D obeys $Ca_c>0.448$, a value notably higher than that observed in 2D. Since identical parameter sets were chosen, the same applies to any other dimensionless number that might control the onset of the defect-rich phase. 
Finally, we note that one recent theory of smectic rheology \cite{Lu08} suggests this is controlled by screw dislocations which have no 2D analogue. Moreover, the residual dislocations seen in the 3D ordered phase have an orientation that also does not exist in 2D. It should be remembered that, in 3D, edge dislocations and screw dislocations are interconvertible: they have the same topological status and the nomenclature merely describes whether the defect line is oriented normal or parallel to the smectic layers \cite{ChaikinLubensky}. The ability of 3D defect lines to reorient as well as translate underscores their greater dynamical freedom compared to similar defects in a 2D geometry. It is thus tempting to attribute the marked difference in dynamics between 2D and 3D systems to the reduced constraints on defect motion that arise in 3D, allowing them to find and annihilate each other more easily.

\section{Acknowledgements}
We acknowledge support by EPSRC Grants No. EP/E045316 and No. EP/E030173, the MAPPER EU-FP7 project (grant no. RI-261507) and computing time at the Argonne Leadership Computing Facility, Argonne, U.S.A. as well as on HECToR, U.K.
We thank A. Morozov for useful discussions. 
M.E.C. holds a Royal Society Research Professorship.

\section{Appendix: Sliding Periodic Boundary Conditions}

The simulations of shear flow are undertaken using a mixed finite
difference / lattice Boltzmann approach, where the former is used
to advance the order parameter in time via the Cahn-Hilliard equation,
and the latter is used to solve the Navier-Stokes equations.
Coupling between the two is
via the velocity field, which advects the order parameter, and
the thermodynamic pressure tensor (6), the divergence of which
contributes a body force on the fluid. The sliding periodic
boundary conditions of Lees and Edwards\cite{leesedwards} may be
applied to the lattice Boltzmann part of the calculation
\cite{Wagner02,Adhikari05} to generate shear in the system. We
describe here how the sliding periodic boundaries are extended to
include the finite difference part of the calculation.

We consider a finite difference mesh where the order parameter
values are co-located with the density and velocity fields provided
by lattice Boltzmann. Away from the sliding periodic boundaries, the
finite difference discretisation is as one would expect.
In order to ensure conservation of order parameter, as expressed by
\begin{equation}
\partial_t \phi + \partial_\alpha (u_\alpha \phi - M\partial_\alpha \mu) = 0,
\end{equation}
we consider each lattice point to be surrounded by a cubic control volume or
cell
of dimensions $\Delta x = 1$ in two or three dimensions.
We compute convective and diffusive fluxes at the cell boundaries using
linear interpolation of the normal velocity component to the cell face.
For the advective fluxes, the interfacial values of the order parameter
are computed via cubic interpolation weighted in the upwind direction.
This is combined with an Euler forward step for the time integration:
a small Courant number ($u\Delta t/\Delta x$), and hence numerically
stability, is guaranteed by the small Mach number constraint
standard in lattice Boltzmann. The gradient of the chemical potential
is computed at each cell face using values of $\mu$ computed at the
lattice point either side.

A body force on the fluid in the Navier-Stokes equations may be
included in the lattice Boltzmann approach. Again, to ensure conservation
of momentum, the divergence of the thermodynamic pressure tensor is
computed based on cell-face values. In practice, this requires a
finite difference stencil
that entends at least $\pm 3$ lattice points in each coordinate direction
(the thermodynamic pressure involves a fourth derivative of the order
parameter).

The sliding periodic boundaries lie (conceptually) along plane
co-incident with the cell faces. Where the finite difference stencil 
overlaps a such a plane, the relative movement between the two sides
of the plane must be accounted for when computing derivatives of the
order parameter, and hence fluxes of order parameter and momentum,
near the planes. This means interpolating values of the order parameter
on each side of the plane to positions corresponding to the lattice
locations on the other side so that the finite difference stencil
can be used normally. For this study, where there are high derivatives
of the order parameter in the free energy, it is important that this
interpolation is at least cubic. Linear interpolation of the order
parameter leads to unphysical oscillations of physical quantities
such as the shear stress related to the magnitude of the velocity jump
across the sliding boundary.
With cubic interpolation, such artefacts are absent. The non-linear
nature of this interpolation means that fluxes of both order parameter
and momentum computed at the plane do not exactly match on each side.
However, this may be corrected at each sliding boundary at each time
step, restoring global order parameter and momentum conservation.

\footnotesize{
\bibliography{smectics} 
\bibliographystyle{rsc} 
}

\end{document}